\documentclass{emulateapj}
\usepackage{apjfonts}
\usepackage{graphicx}
\usepackage{amsmath}

\newcommand{\lsim}{\lower0.6ex\vbox{\hbox{$ \buildrel{\textstyle <}\over{\sim}\ $}}}
\newcommand{\gsim}{\lower0.6ex\vbox{\hbox{$ \buildrel{\textstyle >}\over{\sim}\ $}}}
\newcommand{\beq}{\begin{equation}}
\newcommand{\eeq}{\end{equation}}

\newcommand{\nsc}{n_{\mathrm{s}}}

\newcommand{\Omegam}{\Omega_{\mathrm{M}}}
\newcommand{\Omegab}{\Omega_{\mathrm{B}}}

\newcommand{\Omegade}{\Omega_{\mathrm{DE}}}

\newcommand{\omegam}{\omega_{\mathrm{M}}}
\newcommand{\omegab}{\omega_{\mathrm{B}}}
\newcommand{\ns}{n_{\mathrm{s}}}
\newcommand{\dr}{\Delta_{\mathcal{R}}^{2}}
\newcommand{\wzero}{w_{0}}
\newcommand{\wa}{w_{\mathrm{a}}}
\newcommand{\apiv}{a_{\mathrm{p}}}
\newcommand{\wpiv}{w_{\mathrm{p}}}

\newcommand{\Pkij}{P_{\kappa}^{\mathrm{ij}}}

\newcommand{\nPkij}{\mathcal{P}_{\kappa}^{\mathrm{ij}}}

\newcommand{\nPkobs}[2]{\bar{\mathcal{P}}_{\kappa}^{\mathrm{#1}\mathrm{#2}}}

\newcommand{\probtot}{P_{\mathrm{tot}}\left(z^{ph} | z\right)}
\newcommand{\probcore}{P_{core}\left(z^{ph}|z\right)}
\newcommand{\probcat}{P_{cat}\left(z^{ph}|z\right)}

\newcommand{\Thetacat}{\Theta\left(\frac{\Delta z_{\mathrm{cat}}}{2} - \left| z - z_{\mathrm{cat}}\right| \right)}

\newcommand{\ellmax}{\ell_{\mathrm{max}}}

\newcommand{\ntomo}{N_{\mathrm{TOM}}}

\newcommand{\fsky}{f_{\mathrm{sky}}}

\newcommand{\dd}{\mathrm{d}}

\newcommand{\gi}{\langle \gamma_{int}^{2} \rangle}

\newcommand{\niz}{n_{i}(z)}
\newcommand{\nspec}{N_{\mathrm{spec}}}
\newcommand{\nspeci}{N_{\mathrm{spec}}^{\mathrm{i}}}

\newcommand{\phzpdf}{P(z^{ph}|z^{sp})}

\newcommand{\zmed}{z_{\mathrm{med}}}

\newcommand{\bit}{\begin{itemize}}
\newcommand{\eit}{\end{itemize}}
\newcommand{\ben}{\begin{enumerate}}
\newcommand{\een}{\end{enumerate}}

\newcommand{\sigmaz}{\sigma_{\mathrm{z}}}
\newcommand{\zbias}{z^{\mathrm{bias}}}

\newcommand{\zphot}{z^{\mathrm{ph}}}
\newcommand{\zcat}{z_{\mathrm{cat}}}
\newcommand{\zphcat}{z_{\mathrm{cat}}^{\mathrm{ph}}}
\newcommand{\Fcat}{F_{\mathrm{cat}}}
\newcommand{\sigcat}{\sigma_{\mathrm{cat}}}
\newcommand{\dzcat}{\Delta z_{\mathrm{cat}}}
\newcommand{\Nacat}{N_{\mathrm{cat}}^{\mathrm{A}}}
\newcommand{\Na}{N^{\mathrm{A}}}

\begin{document}
\submitted{The Astrophysical Journal, submitted}
\vspace{1mm}

\shortauthors{Hearin et al.}
\shorttitle{Catastrophic Photo-z Errors and Weak Lensing Tomography}

\title{
A General Study of The Influence of Catastrophic Photometric Redshift Errors on Cosmology with Cosmic Shear Tomography
}

%
\author{
Andrew P. Hearin\altaffilmark{1}, Andrew R. Zentner\altaffilmark{1}, Zhaoming Ma\altaffilmark{2,3}, and Dragan Huterer\altaffilmark{4}
}

\vspace{2mm}


\begin{abstract}

A goal of forthcoming imaging surveys is to use weak gravitational lensing shear measurements 
to constrain dark energy.  A challenge to this program is that redshifts to the lensed, 
source galaxies must be determined using photometric, rather than spectroscopic, information.  
We quantify the importance of uncalibrated photometric redshift outliers to the dark energy 
goals of forthcoming imaging surveys in a manner that does not assume any particular 
photometric redshift technique or template.  In so doing, we provide an approximate blueprint 
for computing the influence of specific outlier populations on dark energy constraints.  
We find that outlier populations whose photo-z distributions are tightly localized about 
a significantly biased redshift must be controlled to a per-galaxy rate
 of $1-3 \times 10^{-3}$ to insure that systematic errors on dark energy 
 parameters are rendered negligible.  In the complementary limit, 
 a subset of imaged galaxies with uncalibrated photometric redshifts distributed 
over a broad range must be limited to fewer than a per-galaxy error rate of 
$F_{\mathrm{cat}} \lesssim 2-4 \times 10^{-4}$.  Additionally, we explore the relative 
importance of calibrating the photo-z's of a {\em core} set of relatively well-understood 
galaxies as compared to the need to identify potential catastrophic photo-z outliers.  
We discuss the degradation of the statistical constraints on dark energy parameters 
induced by excising source galaxies at high- and low-photometric redshifts, 
concluding that removing galaxies with photometric redshifts $\zphot \gtrsim 2.4$ 
and $\zphot \lesssim 0.3$ may mitigate damaging catastrophic redshift outliers at a 
relatively small ($\lesssim 20\%$) cost in statistical error.  In an appendix, we show that 
forecasts for the degradation in dark energy parameter constraints due to uncertain photometric 
redshifts depend sensitively on the treatment of the nonlinear matter power spectrum.
In particular, previous work using PD96 may have overestimated the photo-z calibration requirements
of future surveys.

\end{abstract}



\altaffiltext{1}{Department of Physics \& Astronomy, The University of Pittsburgh, Pittsburgh, PA 15260}
\altaffiltext{2}{Brookhaven National Laboratory, Upton, NY 11973}
\altaffiltext{3}{Department of Physics \& Astronomy, University of Pennsylvania, Philadelphia, PA 19104}
\altaffiltext{4}{Physics Department, University of Michigan, Ann Arbor, MI 48109}
%

\section{INTRODUCTION}
\label{section:introduction}

Weak gravitational lensing of galaxies by large-scale structure is developing into a powerful cosmological probe 
\citep[e.g.,][]{hoekstra_etal02,pen_etal03,jarvis_etal03,van_waerbeke_etal05,jarvis_etal06,semboloni_etal06,kitching_etal07,benjamin_etal07,dore_etal07,fu_etal08}.  
Forthcoming imaging surveys such as the Dark Energy Survey (DES), 
the Large Synoptic Survey Telescope (LSST), the European Space Agency's Euclid, 
and the Joint Dark Energy Mission (JDEM) expect to exploit measurements of weak gravitational 
lensing of distant source galaxies as one of the most effective means to 
constrain the properties of the dark energy 
\citep[e.g.,][]{hu_tegmark99,hu99,huterer02,heavens03,refregier03,refregier_etal04,
song_knox04,takada_jain04,takada_white04,dodelson_zhang05,ishak05,albrecht_etal06,
zhan06,munshi_etal08,hoekstra_jain08,zentner_etal08,zhao_etal09}.  
The most stringent dark energy constraints can be achieved when source 
galaxies can be binned according to their redshifts, yielding a tomographic 
view of the lensing signal.  Among the contributions to the dark energy error budget will be 
the error induced by the need to use approximate redshifts determined from photometric data 
\citep{bolzonella_etal00,collister_lahav04,feldmann_etal06,banerji_etal08,
brammer_etal08,oyaizu_etal07,lima_etal08,dahlen_etal08,abdalla_etal08,newman08,
ilbert_etal09,coupon_etal09,cunha_etal09,schulz09}
because it is not possible to obtain spectroscopic redshifts 
for the large numbers of source galaxies needed to trace cosmic shear.  
Photometric redshifts are and will be calibrated by smaller samples of 
galaxies with spectroscopic redshifts.  In this paper, we study the influence of 
poorly-calibrated photometric redshifts for small subsets of the galaxies 
within imaging samples on dark energy constraints.

The influence of uncertain photometric redshifts (photo-z hereafter) 
on the dark energy program has been studied by a number of authors 
\citep{ma_etal06,huterer_etal06,lima_hu07,kitching_etal08,ma_bernstein08,sun_etal09,
zentner_bhattacharya09,bernstein_huterer09,zhang_etal09}.    
Studies of the requirements for photo-z accuracy have assumed relatively 
simple forms for the relationship between the inferred photo-z of a galaxy and its spectroscopic redshift, in particular
that this is a Gaussian distribution with a redshift-dependent bias and scatter.  The underlying 
assumption is that this distribution can be calibrated with an appropriate spectroscopic 
sample over the range of redshifts of interest.  These studies indicate that roughly 
$\nspec \sim 10^5$ spectroscopic redshifts are needed to render photo-z uncertainty 
a small contributor to the dark energy error budget, but any particular number depends 
upon the many details of each study.  Broadening the description of the photo-z distribution 
to a multi-component Gaussian leads to slightly more demanding requirements on the 
spectroscopic calibration sample \citep{ma_bernstein08}.  
However, complexity or multi-modality of the photo-z distribution will not induce large systematic 
errors on dark energy parameters, provided that this complexity 
is known and that we have some ability to calibrate complex photo-z 
features using spectroscopic galaxy samples (a non-trivial assumption).  
That is not to say that dark energy constraints are insensitive to such complexity.  
Broad or multi-modal photo-z distributions will provide an effective limit to the redshift 
resolution of tomographic weak lensing and will degrade dark energy constraints.  If this complexity 
can be diagnosed in spectroscopic samples, it may be treated by generalizing the modeling 
in \citet{ma_etal06} and \citet{ma_bernstein08}.  In approximate accordance with the prevailing 
nomenclature, we refer to the galaxies for which spectroscopic calibration of the photo-z distribution 
will be possible as the {\em core} photo-z distribution.  \citet{ma_bernstein08} studied multi-modal 
core photo-z distributions in some detail.

These studies assume that the spectroscopic samples that will be obtained will suffice 
to calibrate the photo-z's of all galaxies utilized in the weak lensing analysis.  However, spectroscopic calibration 
samples may well be deficient in spectra of some subset of galaxies that otherwise may not be easily 
identified and removed from the imaging sample \citep[for example, see][for a discussion]{newman08}.  
Consequently, some fraction of galaxies in forthcoming imaging samples may not have photo-z's that are well 
calibrated spectroscopically and may have photo-z's that differ markedly from their true redshifts.  
Including such galaxies in weak lensing analyses would lead one to infer biased estimators of dark energy parameters.
  These systematic offsets in dark energy parameters may be considerable compared to statistical errors.

We refer to  such subsets of galaxies that are not well calibrated by spectroscopic samples and 
which have photo-z distributions that differ markedly from the photo-z distributions of the core galaxy samples 
as {\em catastrophic} photo-z outliers.  Our chief aim in this study is to estimate the biases induced 
on dark energy estimators by catastrophic photo-z outliers for a variety of possible manifestations of catastrophic outliers, and 
to estimate the level at which such outliers must be controlled in order to mitigate dark energy biases.

We consider two broad classes of catastrophic outliers, differentiated by the breadth of their photo-z distributions.  
We emphasize that our definition of a catastrophic outlier is more inclusive than previous usage \citep[compare to][]{bernstein_huterer09}.
Outlier populations with photo-z's that are confined to a small range of highly-biased redshifts make up the class we refer to as 
\textit{localized catastrophes}.  As an example, such outliers may correspond to galaxy populations in which spectral features have been 
misidentified in broadband photometric observations; the prevailing usage of the term \textit{catastrophic error} closely resembles 
our usage of the term \textit{localized catastrophe}.  
The second class of outliers we consider, which we refer to as \textit{uniform catastrophes}, have photometric redshifts that are relatively unconstrained.  
This class may more naturally be associated with a level of spectroscopic incompleteness yielding a population of imaged galaxies with 
little information on the reliability of their photometric redshifts.  

We describe our modeling techniques in \S~\ref{section:methods}.  
We detail our results on the potential importance of catastrophic 
photo-z outliers in \S~\ref{section:results}.  This section includes a 
brief discussion of mitigation strategies in which we explore the 
possibility of eliminating subsets of galaxies in order to reduce 
biases at the cost of increased statistical errors.  We discuss the implications 
of our results in \S~\ref{section:discussion} and 
summarize our work in \S~\ref{section:summary}.

We include in this study an appendix that may help in comparing published results 
on photometric redshift calibration requirements.  
All treatments of dark energy constraints from weak lensing rely on some approximate 
treatment of the growth of structure in the nonlinear regime.  Several approaches 
are in common use 
\citep{scherrer_bertschinger91,peacock_dodds96,seljak00,ma_fry00,scoccimarro_etal01,cooray_sheth02,smith_etal03} 
and additional parameters have been introduced to model baryonic processes 
\citep{rudd_etal08,zentner_etal08,guillet_etal09}.  
In the main body of our paper, we use the fitting form provided by \citet{smith_etal03}.  
We demonstrate in the appendix that estimates of photo-z calibration requirements 
depend upon the modeling of nonlinear power.  Implementing the 
\citet{smith_etal03} relation for nonlinear power results in significantly 
reduced photo-z calibration requirements as compared to previous 
results \citep[e.g.,][]{ma_etal06} that employed the 
\citet{peacock_dodds96} approximation.

\section{Methods}
\label{section:methods}

In this section we describe the methods used in our analysis.  We begin in \S~\ref{sub:photoz} 
with a discussion of our treatment of photometric redshifts, including both the 
{\em core} photometric redshift distributions as well as 
{\em catastrophic} outliers.  
In \S~\ref{sub:wl}, we describe our weak lensing power spectrum 
observables.  We describe cosmological parameter forecasting in 
\S~\ref{sub:Fisher} and conclude with a description of our fiducial 
cosmology and representative surveys in \S~\ref{sub:model}.

\subsection{Photometric Redshift Distributions of Source Galaxies}
\label{sub:photoz}

We characterize the distribution of photometric redshifts through the probability of obtaining a photometric redshift $\zphot$, given a galaxy with 
spectroscopic (or "true") redshift $z$, $P(\zphot | z)$.  The distribution of true redshifts of galaxies in a photometric bin labeled with index 
$i$ is
\beq
\label{eq:dnidz}
\niz=n(z)\int_{z_{i}^{\mathrm{low}}}^{z_{i}^{\mathrm{high}}} \ \dd z^{ph}\ P(\zphot | z),
\eeq
where $n(z)$ is the number density of source galaxies per unit redshift $z$, 
$\niz$ is the number density of sources per unit redshift that are assigned to 
the $i^{\mathrm{th}}$ photo-z bin, and $z_{i}^{\mathrm{low}}$ and $z_{i}^{\mathrm{high}}$ 
delineate the boundaries of the $i^{\mathrm{th}}$ tomographic bin.  

We model the overall galaxy distribution via 
\beq
\label{eq:galaxydistribution}
n(z) \propto z^{2}\ \exp[-(z/z_{0})^{1.2}],
\eeq
where $z_{0}$ is determined by specifying the median redshift of the survey and the powers of 
redshift are representative of the distributions of observed high-redshift galaxies \citep{newman_etal10}.  
The normalization of the overall galaxy distribution is determined by the total 
number of galaxies per unit solid angle, 
$$N^{A}=\int_{0}^{\infty}\ \dd\ z\ n(z),$$ 
and we designate the number of galaxies per solid angle 
in any photo-z bin as 
$$N_{\mathrm{i}}^{A}=\int_{0}^{\infty}\ \dd\ z\ n_{i}(z) .$$
%

\subsubsection{The Core Photometric Redshift Distribution}
\label{sub:core}

For the purposes of our study, we consider the {\em core} galaxy distribution to be 
comprised of galaxies with a {\em photometric} redshift distribution that will be well characterized 
through calibration with spectroscopically-observed galaxy samples.  
Studies using existing spectroscopic galaxy samples to predict the photo-z distributions of galaxies 
in future large-scale image surveys indicate that the core distributions may be complicated 
\cite[e.g.,][]{jouvel_etal09,ilbert_etal09,coupon_etal09}.  
A common simplifying assumption in the literature is that the photometric redshifts of galaxies in the 
core are distributed according to a Gaussian distribution with a redshift-dependent mean and 
variance \citep[e.g.,][]{ma_etal06,ma_bernstein08}, 
 \beq
\label{eq:pcore} 
P_{core}(z^{ph}|z)=\frac{1}{\sqrt{2\pi}\sigma_z}\exp\Bigg[-\frac{(z-z^{ph}-z^{bias})^2}{2\sigma^2_z}\Bigg],
\eeq
where both $\sigma_{z}(z)$ and $z^{\mathrm{bias}}(z)$ are functions of true redshift, $z$.  
The redshift-dependent mean and variance endow this form with sufficient flexibility to treat a 
wide variety of redshift distributions; however, this simple model does neglect complex features 
that may be present in the realized photometric redshift distributions of future surveys.  
We adopt this model because it is a published standard against which our results can be 
compared, and because the complexity of calibrating the core sample of photometric 
redshifts is not the primary aim of our work.  

We compute the functions $\sigma_{z}(z)$ and $z^{\mathrm{bias}}(z)$ by linear interpolation between 
values tabulated at 31 redshift points spaced evenly between $z=0$ and 
$z=3$.  This choice of binning allows for maximal degradation in dark energy constraints absent 
prior information about the photometric redshift distribution of source galaxies.  
We treat the bias and dispersion at each of these redshifts as free 
parameters in our forecasts, so that there are $2 \times 31 = 62$ 
free parameters describing the core photometric redshift distribution.  
For our fiducial model, we take $\sigma_z(z)=0.05(1+z)$ and 
$z^{\mathrm{bias}}(z)=0$.

\subsubsection{Catastrophic Photometric Redshift Outliers}
\label{sub:cat}

Forthcoming large imaging surveys will observe a tremendous number of galaxies.  
It is unlikely that accurate calibration of every class of photometric redshift distribution 
will be made, at least in part due to the limitations of obtaining reliable spectroscopic redshifts 
\citep[e.g.,][]{newman08} and observations of relatively rare objects.  If either the uncalibrated objects 
follow the redshift distributions of the sample of calibrated photometric redshifts, or the uncalibrated 
objects can be identified from imaging data and removed from the sample, they will have a 
relatively benign impact on the dark energy aims of these surveys.  In the former case, they 
present no systematic error because they follow the redshift distribution of the majority of galaxies, 
and in the latter case they can be removed from the imaging sample at a small cost in statistical 
uncertainty.  Conversely, if a sample of uncalibrated source galaxies that does not follow 
the redshift distribution of the calibrated sources remains in the imaging data used for dark energy 
constraints, this could represent a significant additional systematic error.  
In approximate accordance with established nomenclature, we refer to subsets 
of galaxies that do not follow calibrated photometric redshift distributions {\it and} 
cannot be removed from imaging data as {\em catastrophic} photometric redshift outliers.

In practice, it is expected that catastrophic photometric redshift outliers will be present at some level in 
forthcoming imaging surveys.  The prevalence of multi-modal features in the photo-z distributions of 
existing calibration samples is a clear illustration of the difficulty of determining galaxy redshifts 
from photometric colors \citep{oyaizu_etal07,cunha_etal09,ilbert_etal09,coupon_etal09}.
When a population of galaxies responsible for a non-trivial photometric redshift determination appears 
sufficiently often in spectroscopic samples, its associated photo-z error can be calibrated, 
perhaps leading to multi-modal features in the core distribution.  However, there will inevitably be 
populations of galaxies with photo-z degeneracies that are sufficiently rare so as to evade spectroscopic sampling,  
the spectroscopic calibration of a truly representative sample will not be complete, and the removal of galaxies 
with troublesome redshifts from the imaging data will be imperfect.  Each of these difficulties leads to a population of 
outlier galaxies, with distributions not described by the core photometric redshift model, that contributes a systematic 
error to dark energy parameter estimators.

\begin{figure}[t]
\centering
\includegraphics[width=8.8cm]{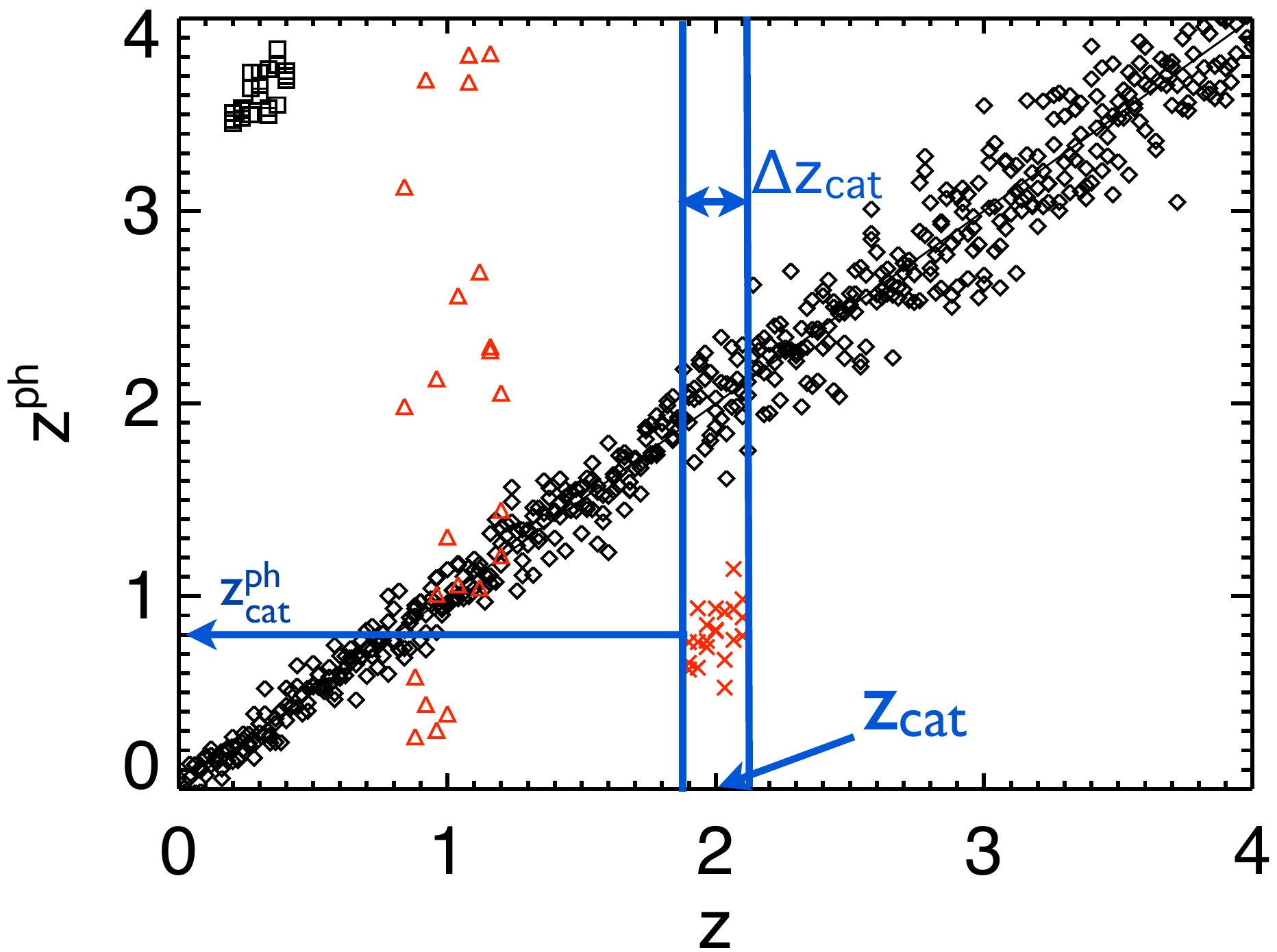}
\caption{ 
A toy illustration of a multi-component photometric redshift distribution.  The aim of this figure 
is to provide a convenient, schematic representation of the photometric redshift distributions we explore.  
{\em Black diamonds} are galaxies in the primary peak of a Gaussian core population of photometric 
redshifts specified by Eq.~(\ref{eq:pcore}).  
{\em Black squares} are galaxies in a secondary peak in a multi-modal core distribution. 
These photometric redshifts are offset from the line $z=\zphot$, but they are a known component of 
the photometric redshift distribution, and if they are represented 
adequately in spectroscopic data they can be calibrated out.  
{\em Red crosses} are galaxies that reside in a catastrophic outlier population with significantly biased, 
but relatively localized, photometric redshifts.  In our nomenclature, this population is 
not represented in spectroscopic calibration samples and contributes a systematic 
error to dark energy parameters.  The {\em red triangles} represent 
galaxies that comprise a uniform catastrophic outlier population, where photometric 
redshifts are relatively unconstrained.  The labels $z_{\mathrm{cat}}$, $\Delta{}z_{\mathrm{cat}}$, 
and $\zphot_{\mathrm{cat}}$ designate the parameters of our catastrophic photometric redshift 
models.
}
\label{fig:toypdf}
\end{figure}

To illustrate the distinction between catastrophic outliers and multi-modal features in the core, 
consider the photo-z distribution illustrated in Figure~\ref{fig:toypdf}.  The bulk of the galaxies in 
this distribution (black diamonds) are scattered about the line $z=\zphot$.  This is a population 
of $400$ galaxies drawn from the Gaussian distribution of 
Eq.~(\ref{eq:pcore}).  There are also two "islands" in the distribution.  The appearance of these island contributions
to the photo-z distribution is quite similar, but they are intended to represent photo-z errors of a qualitatively different nature,
as discussed below.  One island has 
$(z,\zphot)$ coordinates $(0.3, 3.7)$ and the other has $(2.0,0.8)$.   The island at 
$(z,\zphot)=(0.3,3.7)$, consisting of black squares, is a schematic representation of some subset of 
galaxies that give a known, calibrated, small probability of yielding a highly-biased photometric redshift.  
This is a component of a multi-modal core distribution and may either be calibrated with spectroscopy 
or removed from the sample.  The island at $(z,\zphot)=(2.0,0.8)$, consisting of the 
red crosses, is a schematic representation of a catastrophic outlier population.  These are a small subset 
of galaxies with true redshifts near $z \approx 2$, that yield strongly biased, but localized, 
photometric redshifts.  Moreover, this is a population that is either not identified and calibrated in 
spectroscopic samples, or is incompletely removed from imaging data, so that this outlier 
contributes a systematic error to the dark energy error budget.  This is the type of error 
that is our focus in this manuscript.  Finally, there is a population of galaxies that 
is localized near $z \approx 1$ and spread uniformly across $\zphot$.  These galaxies 
represent another extreme of catastrophic photo-z errors in that the redshifts may not be 
strongly biased, but they are poorly constrained and will contribute systematic 
errors for dark energy.

We emphasize the distinction between $\phzpdf$ and the posterior redshift distribution for 
an individual galaxy resulting from a photometric redshift estimation algorithm, often denoted as $p(\zphot)$.  
Often, a single redshift 
estimate is assigned to a galaxy.  In this case, each point in Fig.~\ref{fig:toypdf} may correspond 
to the true redshift and the estimated redshift of a galaxy or population of galaxies.  One may 
utilize more of the information in the $p(\zphot)$, in which case the local density of 
points in Fig.~\ref{fig:toypdf} may correspond to regions in which the posterior has non-negligible 
support.  Our aim is to outline a set of general impacts induced by making large, uncalibrated 
photometric redshift errors.  We use $\phzpdf$ to quantify these effects because this allows for 
a very general characterization of the influences of photometric redshift errors (or equally, 
errors in compressing the information contained in the posteriors).  This is sensible because 
$\phzpdf$ can be constructed from the posteriors of a calibration set in a straightforward manner, 
but this relationship is not invertible so that general statements are difficult or impossible to make.  

\subsubsection{Localized Catastrophic Outliers}
\label{sub:localcat}

One cause for localized catastrophic redshift errors (such as the {\em red crosses} in Fig.~\ref{fig:toypdf}) 
is the misidentification of a spectral feature in broadband photometric observations of galaxies over some range of true redshift.  
A specific example of this occurs when the Lyman-break is confused with the 
$4000\ \mathrm{\overcirc{A}}$-break.  The effect on the photo-z distribution of a small portion of errors due to 
Lyman-$4000\ \mathrm{\overcirc{A}}$ confusion would look something like the small island of squares at 
$(z , \zphot) = (0.3, 3.7)$ in Fig.~\ref{fig:toypdf} \citep{bernstein_huterer09}.  Confusion between the Lyman and 
$4000 \mathrm{\overcirc{A}}$ breaks may occur often enough in spectroscopic samples to be calibrated and thus included as 
a secondary peak in the core distribution, but it is possible that there will be other small 
redshift windows where tertiary islands remain uncalibrated.  

Throughout this paper we adopt a simple model for the photo-z distributions of 
{\em localized catastrophes} as Gaussians with spreads $\sigma_{cat}$ centered 
away from the core at $\zphot_{\mathrm{cat}}$, 
\beq
\label{eq:pgausscat}
P_{\mathrm{cat}}(\zphot | z) = \frac{1}{\sqrt{2\pi}\sigma_{cat}} \exp \left[-\frac{(\zphot - \zphot_{\mathrm{cat}})^2}{2\sigma_{cat}^2} \right].
\eeq
The parameter $\zphot_{\mathrm{cat}}$ specifies the location of the island in photometric redshift, 
and $\sigma_{\mathrm{cat}}$ gives the spread of the catastrophe in $\zphot$.  
In the presence of a localized catastrophe the total photometric redshift 
distribution is 
\begin{eqnarray}
  \probtot & =  & \left[1-\Xi(z)F_{cat}\right] \probcore \nonumber \\ 
                 & +  &  \Xi(z)F_{cat}\probcat.
\end{eqnarray}
The catastrophic error 
occurs over only a specified range of true redshifts, 
$z_{\mathrm{cat}} - \Delta z_{\mathrm{cat}}/2 < z < z_{\mathrm{cat}} + \Delta z_{\mathrm{cat}}/2$, 
as enforced by the function
\beq
\label{eq:thetacat}
\Xi(z) \equiv \Thetacat,
\eeq
where $\Theta(x)$ is the Heaviside step function.  
The quantities $\zphot_{\mathrm{cat}}$ 
(location of the local catastrophe in photometric redshift), 
$z_{\mathrm{cat}}$ (central value 
of the range of true redshifts over which the catastrophe occurs), 
$\Delta z_{\mathrm{cat}}$ 
(width of the range of true redshifts over which the catastrophic error is made), and 
$\sigma_{\mathrm{cat}}$ (spread in $z^{\mathrm{ph}}$ of the catastrophe) 
are four of the five parameters that specify the local catastrophe model.  
The fifth parameter, $F_{\mathrm{cat}},$ is the fraction of galaxies in the true redshift window set 
by $z_{\mathrm{cat}}$ and $\Delta z_{\mathrm{cat}}$ for which the catastrophic error occurs.  

The term $\Xi(z)F_{\mathrm{cat}}$ removes the appropriate fraction of galaxies 
from the core distribution and ensures that $\int_0^{\infty}\ \dd \zphot \ \probtot = 1$.  
As a concrete example, the catastrophic outliers represented by the red crosses in 
Fig.~\ref{fig:toypdf} are galaxies drawn from our model with $F_{\mathrm{cat}}=0.03$, 
$\zphot_{\mathrm{cat}}=0.8$, $z_{\mathrm{cat}}=2.0$, $\sigma_{\mathrm{cat}}=0.1$, 
and $\Delta z_{\mathrm{cat}}=0.1$. For the sake of pragmatism, we present results for 
localized catastrophes in interesting limits of this five-dimensional parameterization rather than 
an exhaustive exploration of these parameters.

\subsubsection{Uniform Catastrophic Outliers}
\label{sub:uniformcat}

Empirically, photometric redshift determination algorithms applied to extant calibration samples  yield 
photometric redshift estimates that are relatively unconstrainted on some subsets of galaxies.  
For example, the photo-z distribution of galaxies in 
both the Canada-France-Hawaii Telescope Legacy Survey (CFHTLS) \citep{coupon_etal09}
 and the Cosmological Evolution Survey (COSMOS) \citep{ilbert_etal09} 
possess such a feature within the range of error rates we explore in this work.  
Unconstrained photometric redshifts represent a regime complementary 
to localized catastrophes.  In this case, photometric redshifts 
may be obtained with nearly equal probability over a significant range of redshift.  
Such broad errors may occur when light from one galaxy is contaminated by light from 
another source nearby in angular separation but at a different redshift.  It is natural 
to expect that such errors will occur most frequently near the peak of the observed galaxy 
number density $n(z)$.

Motivated by the presence of such errors, we also treat the extreme case of relatively unconstrained photometric redshifts 
by using a uniform distribution for $\zphot,$ over a symmetric window in 
true redshift centered on $z_{\mathrm{cat}}$ and spanning a width of 
$\Delta z_{\mathrm{cat}}$.  We refer to this kind of error as a {\em uniform} catastrophe for simplicity.  
In the presence of a uniform catastrophe the total photometric redshift 
distribution is
\begin{eqnarray}
\label{eq:pflatcat}
  \probtot & = & \left[1-\Xi(z)F_{\mathrm{cat}}\right] \probcore  \nonumber \\
  & + & \Xi(z) F_{\mathrm{cat}}/(z_{\mathrm{max}} - z_{\mathrm{min}}),
\end{eqnarray}
 where $z_{\mathrm{min}}$ and $z_{\mathrm{max}}$ delineate the photometric redshift
 range of the survey.
In analogy to localized catastrophes, the function $\Xi(z)$ restricts the true redshift range 
over which flat catastrophes occur and $F_{cat}$ specifies the fraction of galaxies in this true 
redshift window whose redshifts are catastrophically in error.  Therefore, three parameters specify this 
simple model, namely $F_{\mathrm{cat}}$, $z_{\mathrm{cat}}$, and $\Delta z_{\mathrm{cat}}$.  
The uniform catastrophe represented by the red triangles in Fig.~\ref{fig:toypdf} is drawn from 
a model with $F_{\mathrm{cat}}=0.05$, $z_{\mathrm{cat}}=1.0$, and 
$\Delta z_{\mathrm{cat}}=0.2$.

\subsection{Cosmic Shear Tomography}
\label{sub:wl}

In this study, we consider constraints from weak gravitational lensing observables only.  
We split source galaxies into $\ntomo$ photometric redshift bins and consider as our 
observables the $\ntomo (\ntomo+1)/2$ distinct number-weighted auto- and cross-power spectra 
of convergence among the source redshift bins.  Unless otherwise stated, we bin source 
galaxies in equal intervals of redshift between $\zphot=0$ and $\zphot=3$ and take 
$\ntomo=5$, resulting in 15 distinct observables.  For this redshift range, five-bin tomography is a useful 
standard because this binning scheme suffices to saturate dark energy constraints 
\citep[][we have verified that this remains so within the parameters of our study as well]{ma_etal06}.

The galaxy number count in each tomographic bin $N_{i}^{A}$, the cross-spectra between 
bins $i$ and $j$, $\Pkij(\ell)$, and the number-weighed spectra $\nPkij(\ell)$ are related by 
\beq
\label{eq:Pkapdef}
\nPkij(\ell) = N_{i}^{A}N_{j}^{A} \Pkij(\ell) = \int_{0}^{\infty}dz\frac{W_{i}(z)W_{j}(z)}{H(z)D_{A}^{2}(z)}P_{\delta}(k=\ell/D_{A},z).  
\eeq
In Eq.~(\ref{eq:Pkapdef}), $\ell$ is the multipole number, $H(z)$ is the Hubble expansion parameter, $D_{A}(z)$ is the angular diameter 
distance to redshift $z$, and 
$P_{\delta}(k,z)$ is the three-dimensional matter power spectrum.  The lensing weight 
functions, $W_{i}(z)$, weight the cosmic shear signal according to the redshift distributions of galaxies within 
each tomographic bin and are defined as 
\beq
\label{eq:lensweightdef}
W_{i}(z)=\frac{3}{2}\Omegam H_{0}^{2}(1+z)D_{A}(z)\int_{z}^{\infty}dz'\frac{D_{A}(z,z')}{D_{A}(z')}n_{i}(z'),
\eeq
where $D_{A}(z,z')$ is the angular diameter distance between redshifts $z$ and $z'$.

\subsection{Parameter Forecasting}
\label{sub:Fisher}

We use the Fisher matrix formalism to study the constraining power of our weak lensing observables 
on dark energy parameters as well as to quantify the systematic errors on dark energy parameters that 
result from catastrophic photometric redshift errors.  
The Fisher matrix formalism is ubiquitous in cosmological parameter forecasting 
\citep[useful references related to the present application include,~][]{jungman_etal96,tegmark_etal97,seljak97,kosowsky_etal02,huterer_takada05,albrecht_etal06,bernstein_huterer09}, 
so we simply quote relevant results here.  The particular implementation we use closely 
mirrors that in \cite{zentner_etal08} and \citet{hearin_zentner09}, to which we refer the 
reader for details.

The Fisher matrix is given by a sum over the observables.  In the particular case of 
weak lensing power spectra, the spectra at different multipoles can be treated as independent 
and this sum can be written as
\beq
\label{eq:fisher}
F_{\alpha \beta}=\sum_{\ell_{\mathrm{min}}}^{\ellmax} (2\ell+1)\fsky 
\sum_{\mathrm{A,B}} \frac{\partial \mathcal{P}_{\mathrm{A}}}{\partial p_{\alpha}} 
[C^{-1}]_{\mathrm{AB}} 
\frac{\partial \mathcal{P}_{\mathrm{B}}}
{\partial p_{\beta}} + F_{\alpha \beta}^{\mathrm{P}},
\eeq
where the $\mathcal{P}_{\mathrm{A}}$ are the set of observables indexed by 
a single label, $\mathbf{C}^{-1}$ is the inverse covariance matrix of these observables at fixed multipole, 
$[C^{-1}]_{\mathrm{AB}}$ are the components of the inverse of the covariance matrix 
(we include the brackets for clarity), and $p_{\alpha}$ are the theoretical model parameters.  
We choose an indexing scheme in which lower-case Greek letters designate model parameters, 
upper-case Latin letters designate observables, and lower-case latin letters 
designate photometric redshift bins, and take the mapping between observable 
number and tomographic bin number to be $A=i(i-1)/2+j$.  
Throughout this paper we use $\ell_{min}=2 \fsky^{-1/2}$, where $\fsky$ is the 
fractional sky coverage of the weak lensing survey, and $\ell_{max}=3000$ as a rough 
indication of the scale beyond which a number of weak lensing approximations 
break down \citep{white_hu00,cooray_hu01,vale_white03,dodelson_etal06,semboloni_etal06,rudd_etal08}.

The covariance matrix of observables at each multipole is
\beq
C_{\mathrm{AB}}(\ell) = \nPkobs{i}{k}(\ell) \nPkobs{j}{l}(\ell) + \nPkobs{i}{l}(\ell) \nPkobs{j}{k}(\ell)
\eeq
where the indices $i$ and $j$ map onto $A$ and $k$ and $l$ map onto $B$.  
The observed number-weighted power spectra, $\nPkobs{i}{j}(\ell)$, have contributions from signal and shot noise, 
\beq
\nPkobs{i}{j}(\ell) = \nPkij(\ell) + N_{i}^{A} \delta_{i j} \langle \gamma_{\mathrm{i}}^2 \rangle ,
\eeq
where the quantity $\langle \gamma_{\mathrm{i}}^2 \rangle$ is the intrinsic source galaxy 
shape noise.  We conform to recent convention and fix $\sqrt{ \langle \gamma_{\mathrm{i}}^2 \rangle }=0.2$, 
so that all deviations from this noise level are incorporated into an effective galaxy number density.

The Fisher matrix formalism provides an estimate of the parameter covariance near a fiducial 
point in the parameter space.  One chooses fiducial values for the model parameters and 
estimates the error on parameter $\alpha$ from the inverse of the Fisher matrix at this point, 
$\sigma({p_{\alpha}}) = [F^{-1}]_{\alpha\alpha}.$  
Within this formalism, statistically-independent prior information about the parameters
 is easily incorporated by simple matrix addition.  The second term in 
Eq.~(\ref{eq:fisher}) is the prior matrix.  In our analysis, we assume {\em independent} 
prior constraints on cosmological parameters, so that the prior matrix reduces to a simple 
diagonal matrix, $F_{\alpha\beta}^{\rm P}=\delta_{\alpha \beta} / (\sigma^{\rm P}_{\alpha})^2$, where 
$\delta_{\alpha \beta}$ is the Kronecker-$\delta$ symbol and $\sigma^{\rm P}_{\alpha}$ is 
the prior 1-$\sigma$, Gaussian constraint on parameter $p_{\alpha}$.  We itemize 
our fiducial model and priors in the following subsection.

Given a systematic error that induces a specific shift in the observables, 
one can use the Fisher matrix to estimate the ensuing systematic error 
in model parameters.  
Using $\Delta \mathcal{P}_{\mathrm{A}}$ to denote the difference between the 
fiducial observables and the observables perturbed by the presence of the 
systematic error, one will infer a set of parameters that is systematically 
offset from the true parameters by 
\beq
\label{eq:bias}
\delta p_{\alpha} = \sum_{\beta} [F^{-1}]_{\alpha \beta} 
\sum_{\ell} (2\ell+1) \fsky \sum_{\mathrm{A,B}} 
\Delta \mathcal{P}_{\mathrm{A}} [C^{-1}]_{\mathrm{AB}} 
\frac{\partial \mathcal{P}_{\mathrm{B}}}{\partial p_{\beta}}.
\eeq
The primary results of our work are estimates of the systematic errors 
in dark energy parameters induced by catastrophic photometric 
redshift outliers.  In related literature, the $\delta p_{\alpha}$ are 
often referred to as {\em biases}; however, we refer to them as 
{\em systematic errors} in order to avoid potential confusion 
with the biases in photometric redshifts.

\subsection{Cosmological Model and Survey Characteristics}
\label{sub:model}

We assume a cosmological model specified by seven parameters.  Three of 
these parameters describe the dark energy.  These three 
parameters are the present energy density in units of the critical density, 
$\Omegade=0.76,$ and two parameters, $\wzero=-1$ and $\wa=0$, that describe a 
linearly-evolving dark energy equation of state, $w(a)=\wzero + (1-a)\wa$ 
\citep[e.g.,][]{linder03,chevallier_polarski, huterer_turner01,albrecht_etal06}.  
The values specified for these parameters are those in our fiducial cosmological model.  
In models with a time-varying dark energy equation of state it is interesting to 
present results for the constraint on $w(a)$ at the scale factor at which it is 
most well constrained.  The scale factor at which $w(a)$ can be best constrained 
is the {\em pivot} scale factor $\apiv$, and is related to the Fisher matrix components as 
\beq
\label{eq;apiv}
\apiv = 1 + \frac{[F^{-1}]_{\wzero \wa}}{[F^{-1}]_{\wa \wa}}.
\eeq
The pivot equation of state parameter is 
\beq
\label{eq:wpiv}
\wpiv \equiv w(\apiv) = \wzero + (1-\apiv)\wa
\eeq
and the error on $\wpiv$ is
\beq
\label{eq:sigmapiv}
\sigma^2(\wpiv) = [F^{-1}]_{\wzero \wzero} - \frac{([F^{-1}]_{\wzero \wa})^{2}}{[F^{-1}]_{\wa \wa}}.
\eeq
The dark energy task force quantifies the constraining power of forthcoming surveys 
according to a figure of merit that reflects the areas of the confidence ellipses 
in the $\wzero$-$\wa$ plane.  In particular, the task force quotes values for the combination 
$\mathcal{F} \equiv [\sigma(\wa) \times \sigma(\wpiv)]^{-1}$ \citep{albrecht_etal06}.

The other cosmological parameters we consider and the fiducial values they assume in our modeling are:  
the non-relativistic matter density $\omegam \equiv \Omegam h^2 = 0.13$; the baryon density 
$\omegab = \Omegab h^2 = 0.0223$; the amplitude of the primordial curvature fluctuations 
$\dr = 2.1 \times 10^{-9}$ (though in practice we vary $\ln \dr$ when computing derivatives of this parameter)
 evaluated at the pivot scale $k_{\mathrm{p}} = 0.05$~Mpc$^{-1}$; 
and the power-law index of the spectrum of primordial density fluctuations $\nsc = 0.96$.  We 
adopt relatively conservative priors of $\sigma^{\rm P}(\omegam)=0.007$, $\sigma^{\rm P}(\omegab)=10^{-3}$, 
$\sigma^{\rm P}(\ln \dr)=0.1$, and $\sigma^{\rm P}(\nsc)=0.04$, each of which is comparable to contemporary, 
marginalized constraints on these parameters \citep{komatsu_etal08}.  
Using marginalized, contemporary priors allows for somewhat more parameter degeneracy 
than may be possible with Planck data and leads to dark energy parameter forecasts that 
are relatively conservative.

In principle, it is relatively straightforward to scale parameter forecasts from one experiment 
to another \citep[e.g.,][]{ma_etal06,bernstein_huterer09}; however, in the interest of simplicity, 
we present explicit results for three specific experimental 
configurations that span the range of observations expected of forthcoming instruments.  

The Dark Energy Survey is the most near-term survey that we consider \footnote{{\tt http://www.darkenergysurvey.org}}.  
We model a DES-like survey by assuming a fractional sky coverage of $\fsky=0.12$ and a surface density 
of imaged galaxies of $N^{A}=15/\mathrm{arcmin}^2$.  
Second, we consider a narrow, deep imaging survey similar      
to a Supernova Acceleration Probe-like implementation of a 
JDEM\footnote{{\tt http://universe.nasa.gov/program/probes/jdem}}$^{\mathrm{,}}$\footnote{{\tt http://snap.lbl.gov/}}.  
We refer to this second type of survey as DEEP and model it with 
$\fsky=0.05$ and $N^{A}=100/\mathrm{arcmin}^2$. 
Lastly, motivated by a future ground-based imaging survey such 
as may be carried out by the LSST\footnote{{\tt http://www.lsst.org}} \citep{LSST}, 
or a space-based mission such as the European Space 
Agency's Euclid\footnote{{\tt http://sci.esa.int/euclid}} \citep{eicbook}, 
we consider a survey with very wide sky coverage taking 
$\fsky=0.5$ and $N^{A}=30/\mathrm{arcmin}^2$.  We refer to this class of survey 
as WIDE.  We assume that the median galaxy redshift in the 
WIDE and DEEP surveys is $z_{\mathrm{med}}=1.0$ and that the median 
galaxy redshift in the DES-like survey is $z_{\mathrm{med}}=0.7$.  
In all cases, we follow recent convention by taking the shape noise 
to be $\sqrt{\gi}=0.2$, subsuming additional noise contributions 
into an effective galaxy number density.  
Table~\ref{table:constraints} summarizes our assumed survey properties.

\section{Results:  Systematic Errors on the Dark Energy Equation of State}
\label{section:results}

In this section, we present the results of our study of catastrophic photometric redshift outliers.  
We begin with the baseline constraints on the dark energy equation of state parameters 
in the limit of perfect knowledge of the source galaxy photometric redshift distribution 
in \S~\ref{sub:baseline}.  We continue in a sequence of increasing complexity.  
We quantify the influence of catastrophic photometric 
redshift errors in the limit of perfect knowledge of the core photometric redshift distribution 
in \S~\ref{sub:knowcore}.  We present results on the influence of catastrophic photometric 
redshift errors in the more realistic case of imperfect knowledge of the core distribution 
in \S~\ref{sub:unccore}.  We explore the prospect of excising galaxies based on their photometric redshifts 
as a simple, first-line defense against systematic errors induced by catastrophic 
photometric redshift errors in \S~\ref{sub:mitigation}.

\subsection{Baseline Constraints}
\label{sub:baseline}


\begin{table}
\caption{Representative Surveys and Baseline Constraints}
\vspace*{-12pt}
\begin{center}
\begin{tabular}{lcccccccr}
\tableline\tableline
\vspace*{-8pt}
\\
\multicolumn{1}{c}{Survey}&
\multicolumn{1}{c}{$\ \ \fsky$} &
\multicolumn{1}{c}{$\ \ \ \ N^{A}\ [\mathrm{arcmin}^{-2}]\ \ $  } &
\multicolumn{1}{c}{$z_{\mathrm{med}}$} &
\multicolumn{1}{c}{$\quad$} &
\multicolumn{1}{c}{$\sigma(\wzero)\ $} &
\multicolumn{1}{c}{$\sigma(\wa)\ $} &
\multicolumn{1}{c}{$\sigma(\wpiv)\ $}&
\multicolumn{1}{c}{$\mathcal{F}$}
\\
\tableline\tableline
\\
DES & $ 0.12 $ & $ 15 $ & $ 0.7 $ &        & $ 0.25 $ & $ 0.77 $  & $ 0.07 $  & $ 18.6 $ \\
WIDE & $ 0.50 $ & $ 30 $ & $ 1.0 $  &      & $ 0.07 $ & $ 0.22 $  & $ 0.02 $  & $ 227.3 $ \\
DEEP & $ 0.05 $ & $ 100 $ & $ 1.0 $  &   & $ 0.10 $ & $ 0.33 $  & $ 0.04 $  & $ 75.6 $ \\
\tableline
\end{tabular}
\end{center}
{\sc Notes.}--- 
Column (1) gives the survey that motivates the particular choice of 
parameters.  Column (2) is the fractional sky coverage of the survey. 
Column (3) gives the effective galaxy number density $N^{A}$, in $\mathrm{arcmin}^{-2}$.  
We have followed current convention and adopted a fixed shape noise of 
$\sqrt{\gi}=0.2$, assuming deviations from this assumption to be encapsulated 
in the effective galaxy number density.  
Column (4) gives the median redshift of galaxies in the survey.  
Columns (5)-(8) give dark energy equation of state constraints in the limit of 
perfect knowledge of the photometric redshift distribution of sources.  These 
include the uncertainty on the pivot equation of state $\sigma(\wpiv)$ and 
the product $\mathcal{F} = [\sigma(\wa) \times \sigma(\wpiv)]^{-1}$.  Note that 
these constraints are from the weak lensing components of these surveys only and 
account for statistical errors only.

\label{table:constraints}
\end{table}


We begin our results section by stating our forecasts for dark energy constraints in the limit of 
perfect knowledge of the photometric redshift distribution.  With little uncertainty in photometric 
redshift distributions, the statistical limits of forthcoming survey instruments would allow for 
constraints on the dark energy equation of state at the level of a few percent, as summarized in 
Table~\ref{table:constraints}.  We emphasize here that the limit of perfect knowledge of the 
photo-z distributions is not the assumption that photometric redshifts are precisely equal 
to the true redshifts of the source galaxies.  Rather, the assumption is that there are 
no catastrophic errors, and that the photometric redshift distribution is described by 
the Gaussian in \S~\ref{sub:core} such that all 62 parameters used to specify the Gaussian 
distribution are known precisely. 

\subsection{Systematic Errors in The Limit of Perfect Core Knowledge}
\label{sub:knowcore}

In this section, we present results for systematic photometric redshift errors in the limit of perfect knowledge of the 
{\em core} distribution of photometric redshifts.  This amounts to the assumption of prior knowledge of the 31 dispersion [$\sigma_z(z)$] 
and 31 bias [$z_{\mathrm{bias}}(z)$] parameters defined in \S~\ref{sub:core} to a level of $ \lesssim 10^{-3}$, which could be 
achieved with a {\em representative} sample of $ \gtrsim 4 \times 10^{5}$ spectroscopic redshifts distributed in redshift in 
a manner similar to those in the imaging survey \citep[see][and the discussion in the appendix of this manuscript]{ma_etal06,ma_bernstein08}.  
This is a simple case to begin with as it allows exploration of the influence of catastrophic redshift errors over a range of 
the catastrophic photo-z parameter space without the additional complications associated with redshift-dependent 
priors on the core photo-z distribution.  This is the limit explored by \citet{bernstein_huterer09}.

\subsubsection{Uniform Catastrophes}
\label{sub:unicat}

First, we address systematic errors induced on dark energy parameters by a small population of 
{\em uniform} catastrophes.  Uniform catastrophes are cases in which some small population of galaxies with true redshifts 
in the range $(z_{\mathrm{cat}}-\Delta z_{\mathrm{cat}}/2 < z < z_{\mathrm{cat}} + \Delta z_{\mathrm{cat}}/2$ yield photometric 
redshift estimates that are distributed broadly in $\zphot$.   This class of error differs from the conventional use of the term 
{\em catastrophic error} and may more naturally be interpreted as a tolerance on spectroscopic incompleteness.  

For simplicity, we take the central redshift of the uniform catastrophe 
to be $z_{\mathrm{cat}}=z_{\mathrm{med}}$, and determine systematic errors as a function of $\Delta z_{\mathrm{cat}},$ the width of the range of redshifts 
over which such errors occur, and $F_{\mathrm{cat}},$  the fraction of galaxies in this range of true redshift that correspond to this type of catastrophic error.  
We refer the reader to Eq.~(\ref{eq:thetacat}) and Eq.~(\ref{eq:pflatcat}) for the expressions that formally define these parameters.  
While we vary these parameters independently, they are both related to the total number density of sources with 
redshifts that are catastrophically in error, 
\beq
\label{eq:ncat}
N^{A}_{\mathrm{cat}}=F_{cat}\int_{z_{\mathrm{cat}}-\frac{\Delta{}z_{\mathrm{cat}}}{2}}^{z_{\mathrm{cat}}+\frac{\Delta{}z_{\mathrm{cat}}}{2}}\ \dd z' \ n(z'), 
\eeq
where $n(z)$ is the overall, true redshift distribution of galaxies.  
We should expect systematic errors to increase with 
both $F_{\mathrm{cat}}$ and $\Delta{}z_{\mathrm{cat}}$ because 
higher values of either parameter result in a greater total number of 
catastrophic errors in the outlier population.

\begin{figure}[t]
\centering
\includegraphics[width=9.0cm]{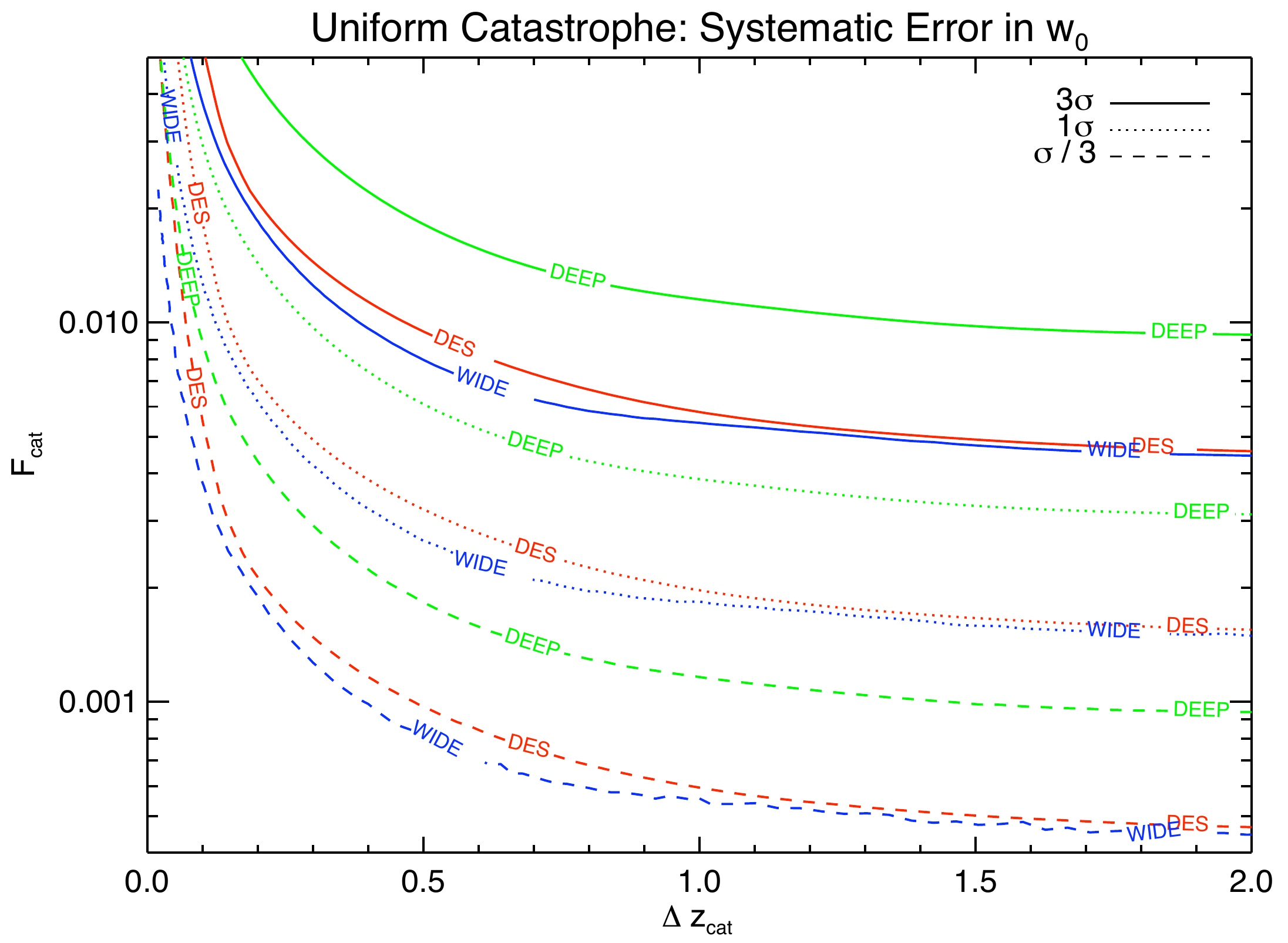}
\includegraphics[width=9.0cm]{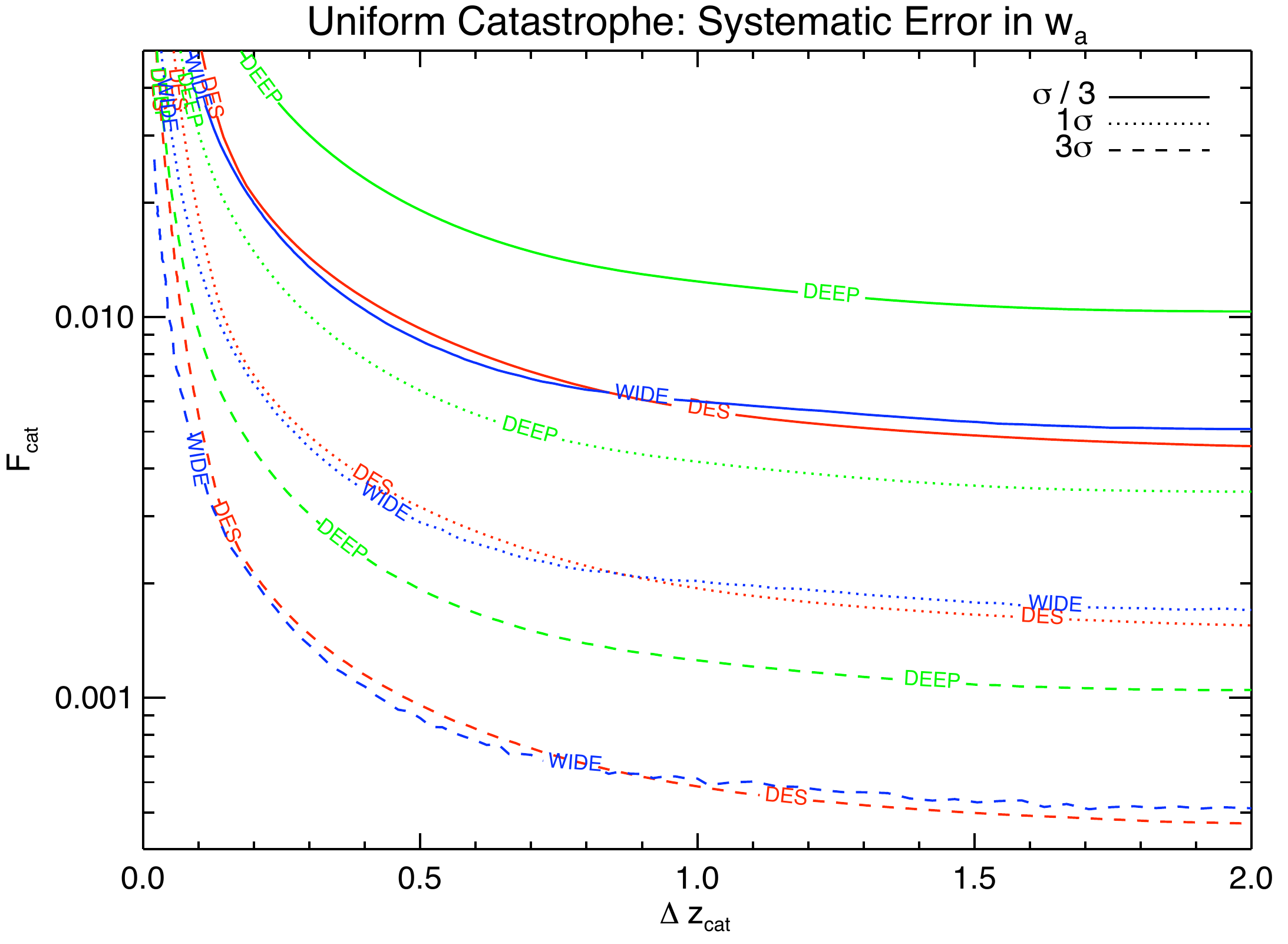}
\caption{
Systematic errors on dark energy parameters in the case of a {\em uniform} photometric 
redshift catastrophe.  The horizontal axis is the width of the range in true redshift over which 
the uniform catastrophe is realized, $\Delta{}z_{\mathrm{cat}}$.  This range in true redshift
is centered at $\zcat=\zmed$ for each experiment 
($\zmed=0.7$ for DES and $\zmed=1$ for DEEP and WIDE).  
The vertical axis is the catastrophic error rate per galaxy within this 
true redshift range, $F_{\mathrm{cat}}$.  The {\em solid} ({\em dashed}) 
lines show contours of constant systematic error equal to three times (one third) the statistical 
errors on each of the dark energy parameters.  The top panel shows contours for 
$\wzero$ and the bottom for $\wa$.  
Chance cancellations in the biases induced by high- and low-redshift galaxies 
cause the DES to be sensitive to catastrophic errors at similar levels to the WIDE 
survey and more sensitive than the DEEP survey.
}
\label{fig:fffwin}
\end{figure}

In Figure~\ref{fig:fffwin} we have quantified the systematic errors induced by uniform catastrophic errors as a function of the 
parameters of our simple model.  The curves in Fig.~\ref{fig:fffwin} are contours of constant systematic error on dark energy 
parameters (for example, $ \vert \delta(\wzero)\vert$ for $\wzero$) 
expressed in units of the statistical error ($\sigma(\wzero)$ for $\wzero$) 
at points in the $\Delta{}z_{\mathrm{cat}}$-$F_{\mathrm{cat}}$ plane.  
For each of the DES, Wide, and Deep surveys, the {\em solid} curves trace 
systematic errors in dark energy that are three times the statistical errors, while the {\em dashed} curves trace 
systematic errors that are 1/3 of the statistical error.  For each of the surveys depicted in Fig.~\ref{fig:fffwin}, 
the region of catastrophic parameter space that is bracketed by the solid and dashed curves labeled with the corresponding 
survey name corresponds to outliers that produce systematic errors which are comparable to statistical errors.  
Systematic errors are relatively small compared to statistical errors in the regions below the dashed 
curves.  Each curve plotted in Fig.~\ref{fig:fffwin} has been generated with a fixed value of $z_{\mathrm{cat}} \equiv \zmed$.  
For the Wide and Deep surveys $\zmed = 1,$ so when $\dzcat = 2$ the uniform catastrophes are made over the true redshift range $0<z<2$.  
For DES $\zmed=0.7,$ so once $\dzcat>1.4$ the true redshift window over which catastrophes are made 
only increases at the high-redshift boundary.

For each of the contours of constant systematic error in Fig.~\ref{fig:fffwin}, 
$F_{\mathrm{cat}}$ decreases with increasing $\Delta{}z_{\mathrm{cat}}$.   
This is simply because increasing the redshift range over which the catastrophic errors are being made ($\Delta{}z_{\mathrm{cat}}$) 
 leads to an increased total number of catastrophic errors, resulting in a decreased tolerance to the error rate, $(F_{\mathrm{cat}}).$
Alternatively, the total number of catastrophic errors in an outlier population
 is given by the integral in Eq.~(\ref{eq:ncat}), and the contours 
of constant systematic error roughly trace constant values of $N^{A}_{\mathrm{cat}}$.  The contours flatten 
considerably for errors that occur over a redshift range $\Delta{}z_{\mathrm{cat}} \gtrsim 0.4$ because there 
will be comparably few imaged sources with true redshifts near $z \sim 0$ or with $z \gtrsim 1.5$.

This treatment of a uniform $\zphot$ catastrophe may appear somewhat contrived but it 
gives insight into a few basic results that are important to recognize.  It is clear that the utility of 
forthcoming shear surveys to constrain dark energy is sensitive to a fractionally small population 
of galaxies that may yield poorly-determined photo-z estimates.  If the error is only relevant to 
galaxies that are relatively isolated in narrow regions of true redshift, for example with 
$\Delta{}z_{\mathrm{cat}} \lesssim 0.1$, then error rates as high as $F_{\mathrm{cat}} \sim 1\%$ 
in this region of true redshift are tolerable.  This is simply because errors that occur with a 
fixed rate over a small redshift range result in a small total 
number of catastrophic outliers to corrupt the weak lensing tomography.  
On the contrary, if such an error occurs for a subset 
of galaxies with true redshifts in an interval of width $\Delta{}z_{\mathrm{cat}} \gtrsim 0.1$, 
then the error rate per galaxy must be significantly lower than $F_{\mathrm{cat}} \lesssim 0.01$ in order 
to render the systematic errors on dark energy equation of state parameters small.

The limit of $\Delta{}z_{\mathrm{cat}} \gg 0.1$ is interesting to consider.  This may correspond 
to the case of a small fraction of galaxies that yield very poorly-constrained photometric redshifts over 
a broad range of true redshifts and that otherwise cannot be identified and removed from the imaging survey.  
In this case, the systematic error from catastrophic photometric redshifts becomes a considerable portion 
of the dark energy error budget at a rate of only $F_{\mathrm{cat}} \sim 10^{-3}$. 
Reducing the systematic error due to such an outlier population to a negligible level requires reducing 
the occurrence of such an outlier population to $F_{\mathrm{cat}} \lesssim 4 \times 10^{-4}$.  Strictly speaking, 
Fig.~\ref{fig:fffwin} corresponds to errors that occur when the true redshift band over which the uniform catastrophe 
occurs is centered on $z_{\mathrm{cat}}=z_{\mathrm{med}}$, but for the $\Delta{}z_{\mathrm{cat}} \gtrsim 1$ 
limit, similar results hold for a wide range of $z_{\mathrm{cat}}$ near unity, 
so this result is of some general relevance to photometric redshift calibration studies.

We conclude this section with a discussion of cancellations that may occur among systematic 
errors.  It may seem somewhat surprising DES exhibits comparable sensitivity to 
uniform errors as WIDE and is more sensitive than DEEP as shown in Fig.~\ref{fig:fffwin}.  
In \S~\ref{sub:local}, we will discuss systematic errors from local catastrophes.  In 
particular, we will show that large biases occur for low $\zcat$ and for higher 
$\zcat$ just over the median redshifts of the surveys (see Fig.~\ref{fig:multicolor}).  
These biases have opposite signs and partially cancel in our forecasts for 
both the DEEP and WIDE surveys.  DES is less sensitive to biases from 
galaxies in the low-redshift range $0.4 < \zcat < 0.6$ that get misplaced to 
higher redshifts because these shifts must compete with the larger 
shot-noise of DES.  The degree of cancellation depends upon modeling choices, 
such as fiducial model and cosmological parameters, but the occurrence of this cancellation 
is robust.  
 
\subsubsection{Localized Catastrophes:  Details}
\label{sub:local}

\begin{figure*}[t]
\centering
\includegraphics[width=5.5cm]{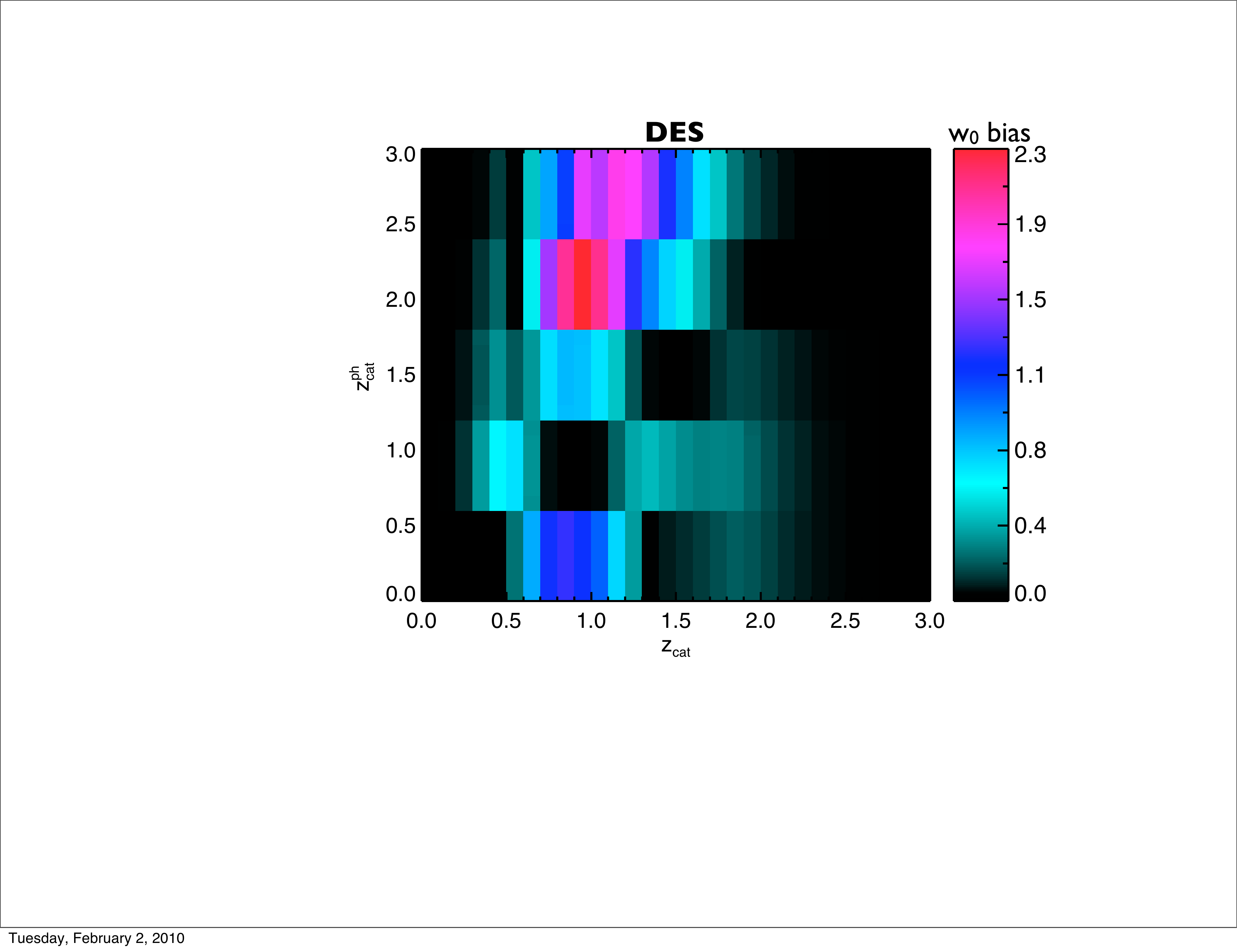}
\includegraphics[width=5.5cm]{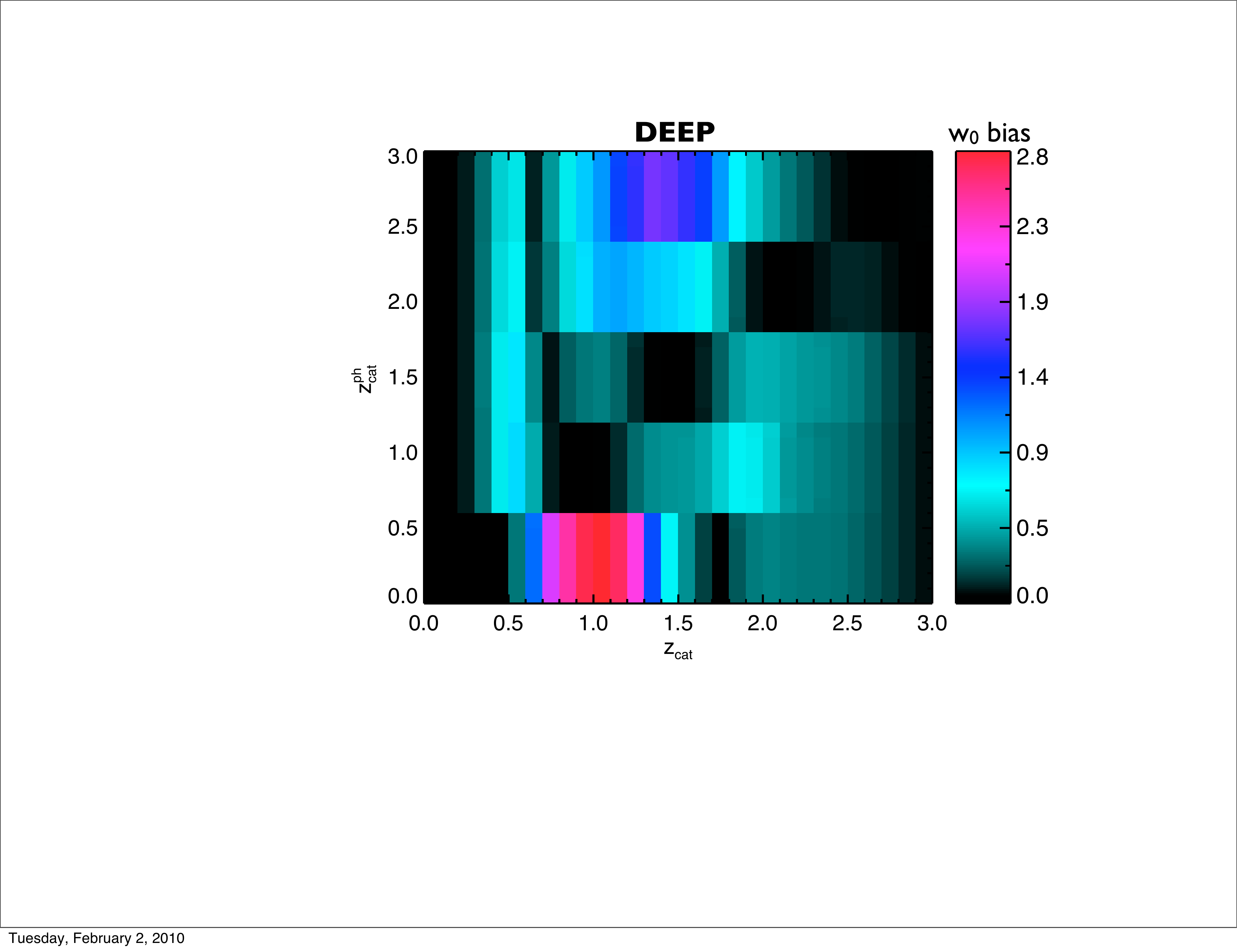}
\includegraphics[width=5.5cm]{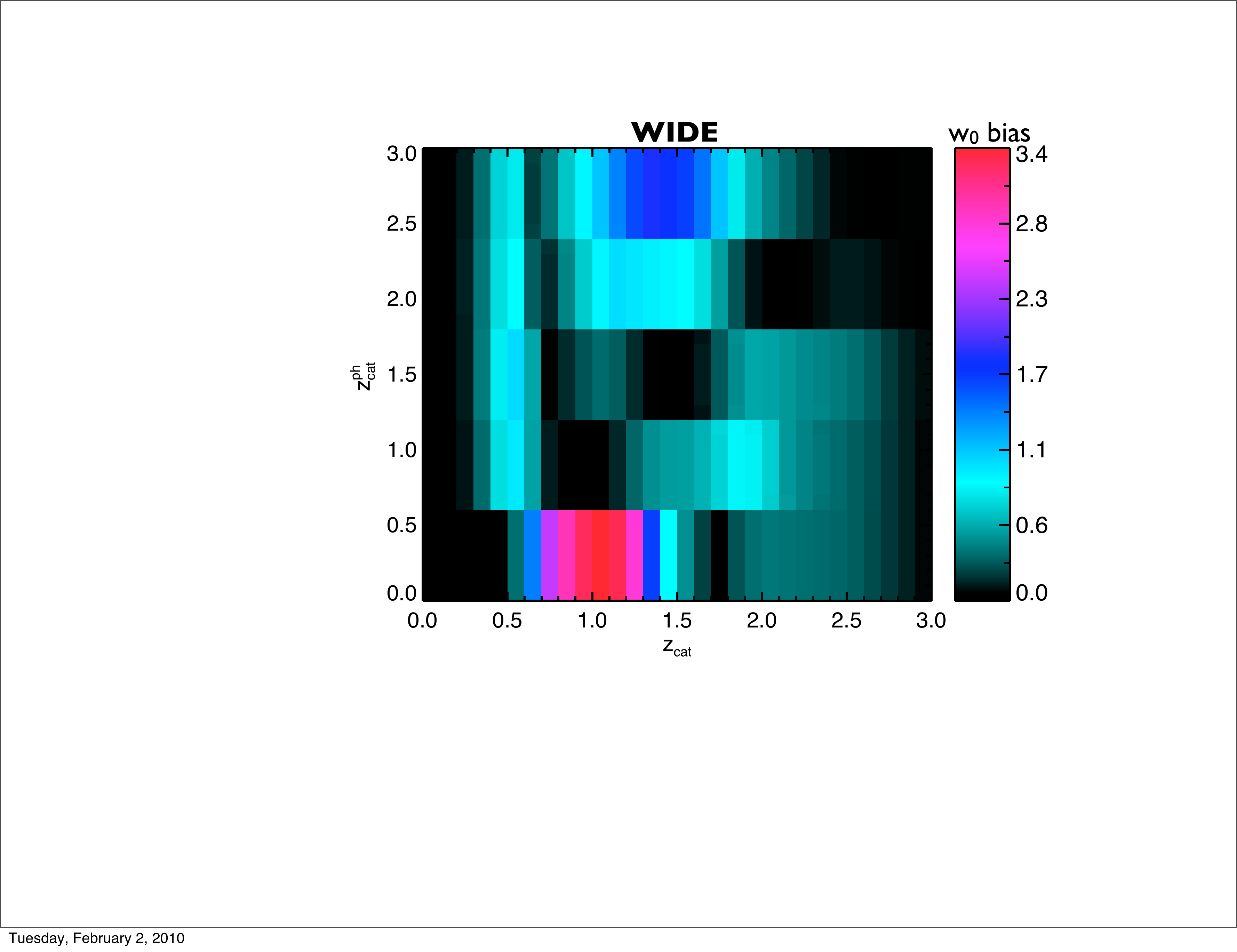}
\includegraphics[width=5.5cm]{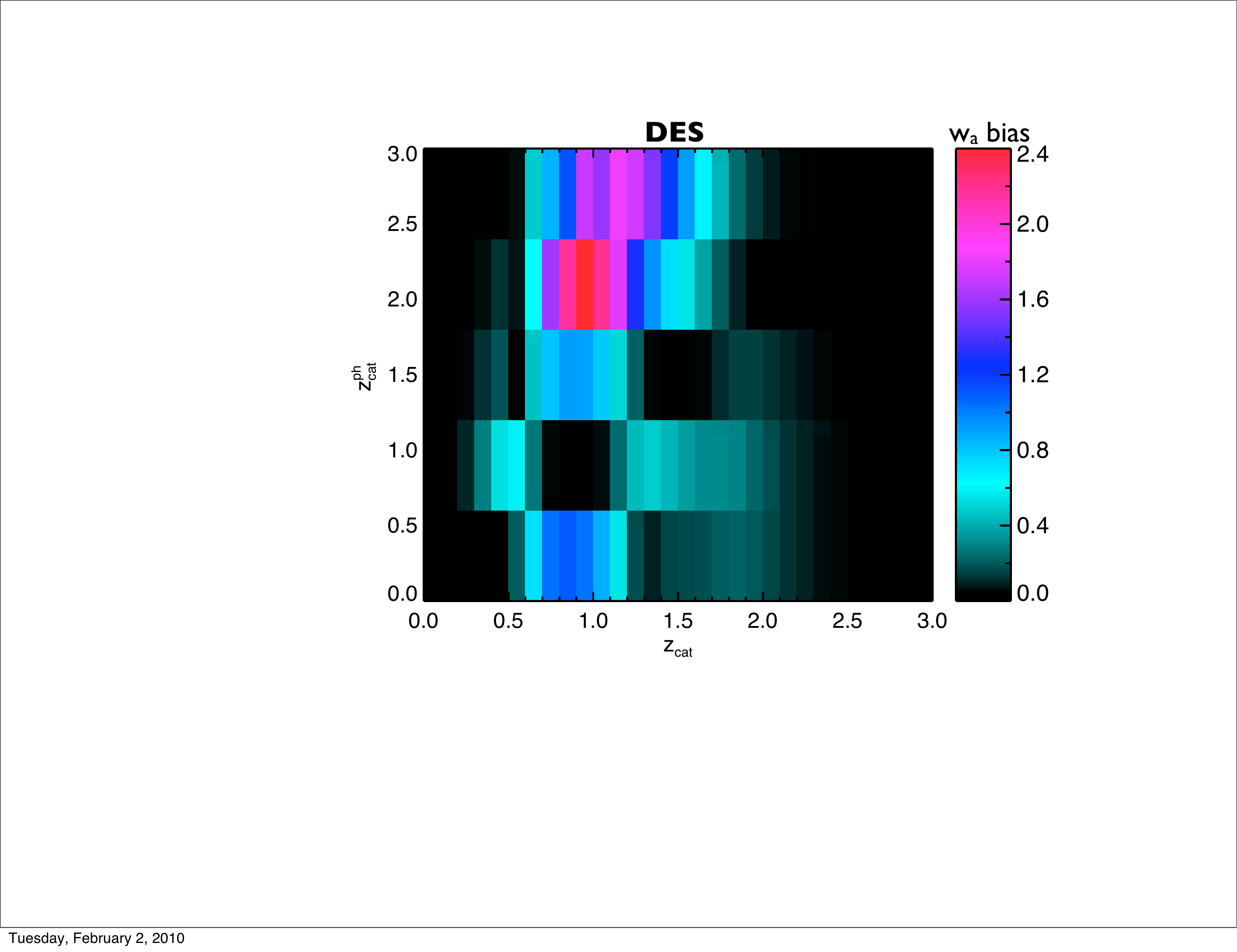}
\includegraphics[width=5.5cm]{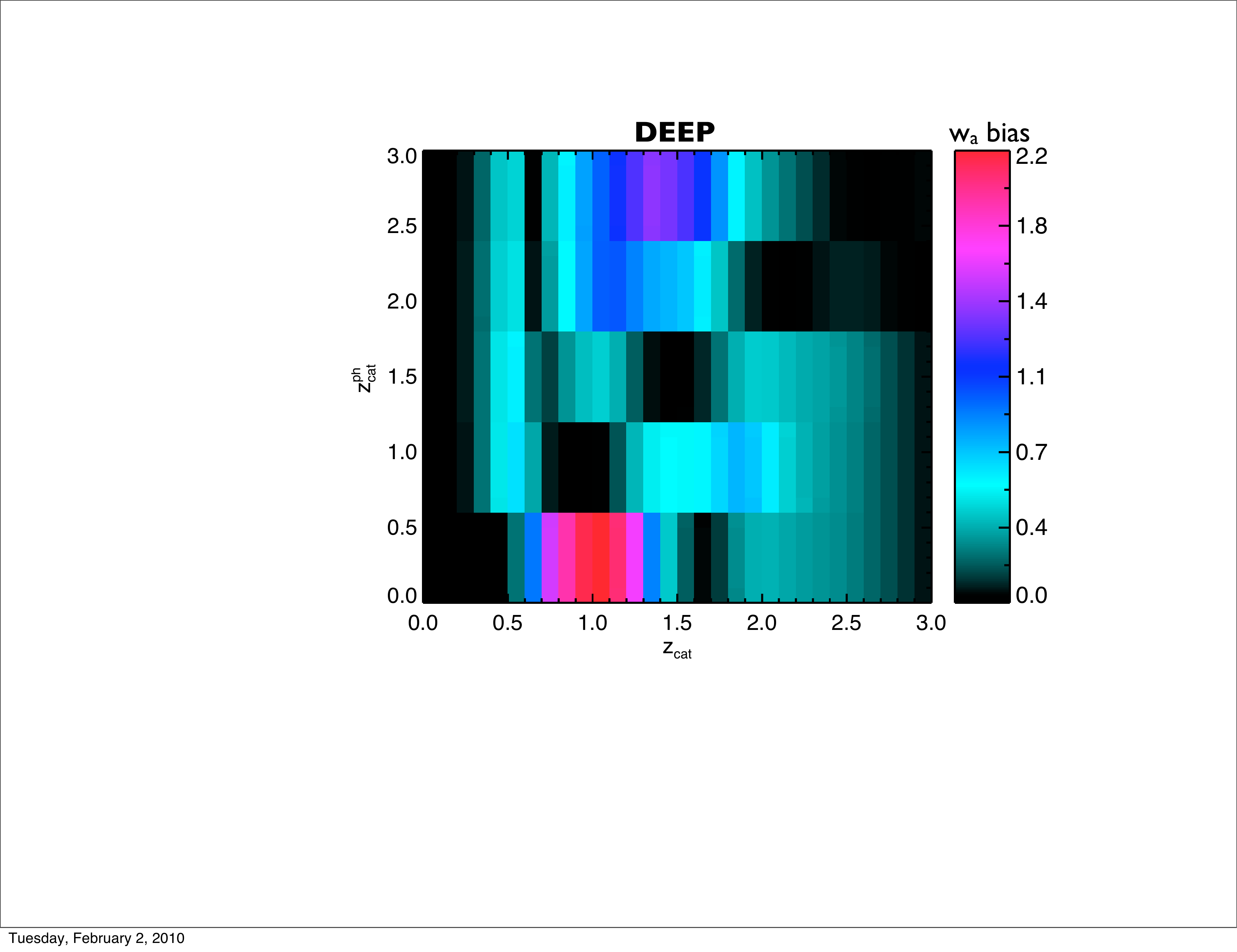}
\includegraphics[width=5.5cm]{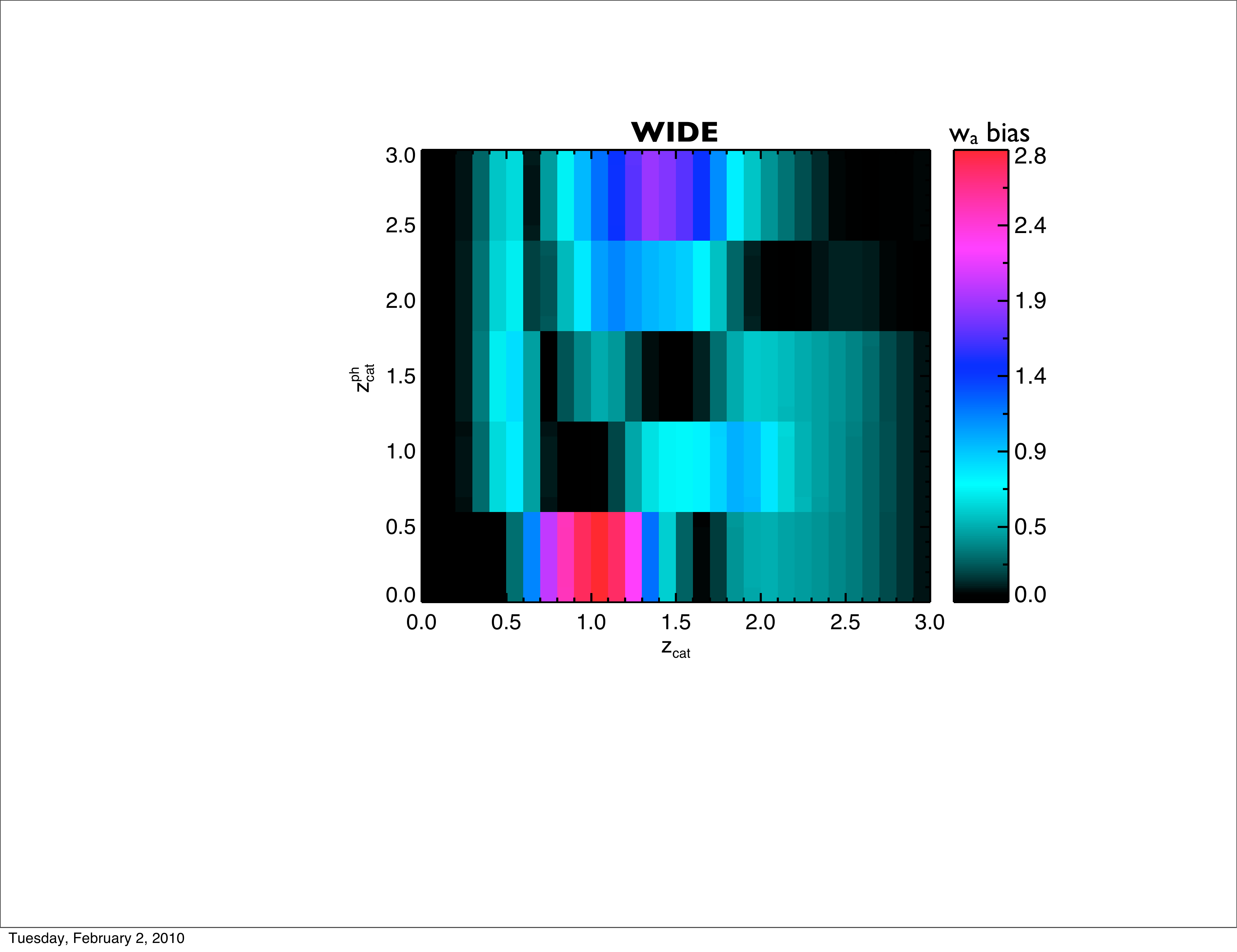}
\caption{
The severity of localized catastrophic errors as a function of the values of source $\zcat$, and target $\zphcat$, of 
the catastrophic errors.  Along the horizontal axes are the values of $\zcat$ while the vertical axes show $\zphcat$, 
just as in Figure \ref{fig:toypdf}.  Each point on this grid corresponds to a localized catastrophe with a fixed per-galaxy 
error rate of $\Fcat=0.05$, and fixed values of both the photo-z spread $\sigcat=0.01$ as well as the width of the 
true redshift range over which the catastrophic error is made, $\dzcat=0.05$.  The effect of these catastrophes on $\wzero$ 
is shown in the top row of panels, while the systematic error on $\wa$ is shown in the bottom row.  
The absolute value of the induced systematic error is color coded; the numerical values labeling the color table 
to the right of each panel indicate the systematic error in units of the statistical uncertainty in the limit of perfect core calibration.
}
\label{fig:multicolor}
\end{figure*}

Localized catastrophes correspond to the case where a small fraction of galaxies near some true redshift $z_{\mathrm{cat}}$ 
yield photometric redshifts that are narrowly distributed about a biased value $\zphot_{\mathrm{cat}}$ that 
is very different from the true redshift $z_{\mathrm{cat}}.$  
Such errors could arise due to incomplete calibration by spectroscopic surveys or 
from difficulty in removing troublesome galaxies from the imaged galaxy sample.  A known example of such 
an error occurs when photo-z algorithms confuse the $4000 \overcirc{\mathrm{A}}$ break with the Lyman break, but other isolated islands of 
biased $\zphot$ persist in contemporary photo-z algorithms (see, e.g. \citealt{coupon_etal09}, \citealt{ilbert_etal09}) 
and may be relevant to forthcoming imaging surveys.

The class of localized photo-z catastrophes is more complex than the uniform case because there are more relevant parameters 
needed to specify the manner in which a localized outlier population is distributed in $\zphot$.   Our toy model requires five parameters 
(see \S~\ref{sub:cat} and Fig.~\ref{fig:toypdf} for an illustration).  
Two are the central value of the true redshift over which this error is operative ($z_{\mathrm{cat}}$) 
and the width of the true redshift range over which this error is operative ($\Delta{}z_{\mathrm{cat}}$).  
Like in the uniform case, some fraction $F_{\mathrm{cat}}$ of galaxies with true redshifts in the 
interval $z_{\mathrm{cat}} - \Delta{}z_{\mathrm{cat}}/2 < z < z_{\mathrm{cat}} + \Delta{}z_{\mathrm{cat}}/2$ 
are catastrophically in error.   The final two parameters specify the biased distribution of photometric redshifts that 
these galaxies are assigned.  These are the (systematically erroneous) value of the photometric 
redshift $\zphot_{\mathrm{cat}}$, and the dispersion in the catastrophic photometric redshift 
distribution $\sigma_{\mathrm{cat}}$ about $\zphot_{\mathrm{cat}}$.

We make an effort to remain agnostic about the classes of photo-z errors that may be realized 
in future imaging data.  However, a complete mapping of even the simple parameter space we 
have specified for catastrophic photo-z's would require a lengthy discussion, so we explore 
useful limits of the model parameters in order to distill our results into a small number of points.    
We are particularly interested in the limit where the source galaxies are placed in a narrow range of 
biased photo-z ($\sigma_{\mathrm{cat}} \ll 0.3$ or so) because the limit of large dispersion in the 
catastrophic photometric redshift population is similar to the {\em uniform} catastrophe of the 
previous section.

We first isolate the sinister regions in the space of $z_{\mathrm{cat}}$-$\zphot_{\mathrm{cat}}$ that lead 
to the most destructive systematic errors in dark energy parameters.  
At a set of points in the parameter space of $(z_{\mathrm{cat}},\zphot_{\mathrm{cat}})$, 
we have calculated the systematic error induced in $\wzero$ and $\wa$ 
by distributing some fraction $F_{\mathrm{cat}}$ of the galaxies with true redshifts near $z_{\mathrm{cat}}$ 
in photometric redshifts centered around some $\zphot_{\mathrm{cat}}$ that is generally very different from $z_{\mathrm{cat}}$.  
We sample a range of values of true redshifts from $\zcat=0.05$ to $\zcat=2.95$, evenly spaced in redshift 
intervals of $\delta{}z=0.1$ and likewise for the photometric redshifts, $\zphcat$.  
In the interest of simplicity, we fix the remaining parameters of our catastrophic photo-z model to 
$F_{\mathrm{cat}} = 0.05$, $\Delta{}z_{\mathrm{cat}}=0.05$, and $\sigma_{\mathrm{cat}}=0.01$ to 
isolate the dependence of the parameter bias upon the location of the catastrophe.

It is important to note explicitly that we present results here at a fixed error fraction $F_{\mathrm{cat}},$
and a fixed true redshift window width and $\Delta{}z_{\mathrm{cat}}.$ 
However, even with these parameters fixed the absolute number of errors varies with $z_{\mathrm{cat}}$  
according to Eq.~(\ref{eq:ncat}), which is roughly 
$N^{A}_{\mathrm{cat}} \sim n(z_{\mathrm{cat}}) \Delta{}z_{\mathrm{cat}}  F_{\mathrm{cat}}$ for 
sufficiently small $\Delta{}z_{\mathrm{cat}}$, along the lines of the analogous discussion for the 
uniform catastrophe in \S~\ref{sub:unicat}.  The aim of this calculation is to map out the relative 
importance of making errors at a {\em fixed rate per galaxy} as a function of the true and photometric 
redshifts of the outliers.

The results of this exercise are depicted in Figure~\ref{fig:multicolor}.  In each column of Fig.~\ref{fig:multicolor} 
there are two panels, corresponding to the systematic errors in $\wzero$ and $\wa$, for each representative experiment.  
The horizontal axes show values of $z_{\mathrm{cat}}$ and the vertical axes 
show values of $\zphot_{\mathrm{cat}}$.  The systematic error is represented on the grid of 
$(z_{\mathrm{cat}},\zphot_{\mathrm{cat}})$ by the color in each of the cells.  In discussing the results of this 
exercise, we find the terminology of \cite{bernstein_huterer09} to be a useful, descriptive shorthand.  
We will refer to the tomographic bin that contains the $z_{\mathrm{cat}}$ value of an outlier as the 
\textit{Source Bin} of that catastrophic photo-z population.  We call the bin containing $\zphot_{\mathrm{cat}}$ its 
\textit{Target Bin}.  This is because galaxies with true redshifts near $z_{\mathrm{cat}}$ are erroneously placed 
in the Target Bin containing the redshift $\zphot_{\mathrm{cat}}$.  Our sampling guarantees that no localized 
outlier straddles a tomographic bin boundary so there are always unique Source and Target bins.  
Outlier populations that straddle a boundary dividing two tomographic bins can be substantially 
more severe than those that do not because such an outlier simultaneously contaminates multiple Target bins.  
We have chosen to ignore such outlier populations for simplicity, but such outliers can be modeled by 
two catastrophic outlier populations, one for each affected target bin.  We will 
return to the issue of tomographic binning and straddling outlier populations below.

The prominent block-like features in Fig.~\ref{fig:multicolor} reflect the tomographic redshift bins 
used in our analysis.  The tomographic bins of the source and target galaxies largely determine both the magnitude 
and sign of the induced systematic error in dark energy parameters.  This gives rise to features that reflect the structure of 
the photometric redshift binning in the $(z_{\mathrm{cat}}$,$\zphot_{\mathrm{cat}})$ plane.  Indeed, for fixed Target and 
Source Bins, the specific value of the target redshift, $\zphot_{\mathrm{cat}}$, within the target photometric redshift bin has little 
influence on the severity of the systematic error.  However, small steps in $\zphot_{\mathrm{cat}}$ can lead to large changes in 
systematic error when the boundary dividing two tomographic bins is crossed.

Varying the location in true redshift, $z_{\mathrm{cat}}$, 
leads to somewhat more significant changes in dark energy systematic error.  
Changing $z_{\mathrm{cat}}$ within fixed Source and Target Bins 
can result in up to a factor of two difference in systematic errors.  
Two factors primarily determine the severity of 
the systematic error as a function of the true redshift of the galaxies, $z_{\mathrm{cat}}$.  
The primary factor stems from the fact that a fixed fractional error rate $(F_{\mathrm{cat}})$
corresponds to a different absolute number of errors $N^{A}_{\mathrm{cat}}$ as a 
function of redshift, $z_{\mathrm{cat}}$.  This is reflected in Eq.~(\ref{eq:ncat}).  
The number of errors $N^{A}_{\mathrm{cat}}$ will be relatively 
large in a region near the median redshift of the survey, 
where the number of source galaxies per unit redshift, $n(z)$, is 
largest.  There are relatively few galaxies at low and high 
redshift, so for a fixed error rate, outlier populations with low or high 
true redshifts contribute a relatively small absolute number of galaxies 
with highly-biased redshifts.

Secondly, an outlier population naturally results in a more severe systematic error the more 
the photometric redshift is biased away from the true galaxy redshift.  Consider the region of catastrophic 
parameter space near $(z_{\mathrm{cat}} , \zphot_{\mathrm{cat}}) = (1.5 , 2.7)$ in either color plot for DES.  
Catastrophes in this region of parameter space correspond to outlier populations whose Source bin 
is the third tomographic bin and Target bin is the fifth tomographic bin.  Outliers in this region of parameter space are 
assigned photo-z's that are significantly too high.  Decreasing the value of $z_{\mathrm{cat}}$ (the "source" of the error) 
increases the distance between core and outlier populations, thereby increasing the systematic error.  This behavior 
contributes significantly to the systematic error gradient near $(z_{\mathrm{cat}},\zphot_{\mathrm{cat}})=(1.5,2.7)$ for 
DES in Fig.~\ref{fig:multicolor}.

Each of the three representative experiments that we consider has two distinct "hot spots" in 
Fig.~\ref{fig:multicolor} that correspond to the most severe types of 
error given a fixed error rate per galaxy, $F_{\mathrm{cat}}$.  A common feature of all these hot spots is their 
$z_{\mathrm{cat}}$ location.  Each of the hot spots lies at a $z_{\mathrm{cat}}$ slightly beyond 
the median survey redshift.  This is sensible because for a fixed error rate, 
the absolute number of catastrophic errors is greatest when they are made at 
the peak in the overall galaxy distribution, that is near  $z_{\mathrm{cat}}=z_{\mathrm{med}}$.  
The most damaging systematic errors occur when the galaxies are shifted to either very low or very high 
photometric redshifts, when the target redshift, $\zphot_{\mathrm{cat}}$, is very different from the source redshift, $z_{\mathrm{cat}}$,   
because the galaxies in error are then placed at distances significantly different from their 
true redshifts.  For our WIDE and DEEP surveys the largest systematic errors tend to occur for galaxies shifted from 
a source redshift $z_{\mathrm{cat}}$ near $z_{\mathrm{med}}=1$ to very low photometric redshifts.  

\begin{figure*}[t]
\centering
\includegraphics[width=8.0cm]{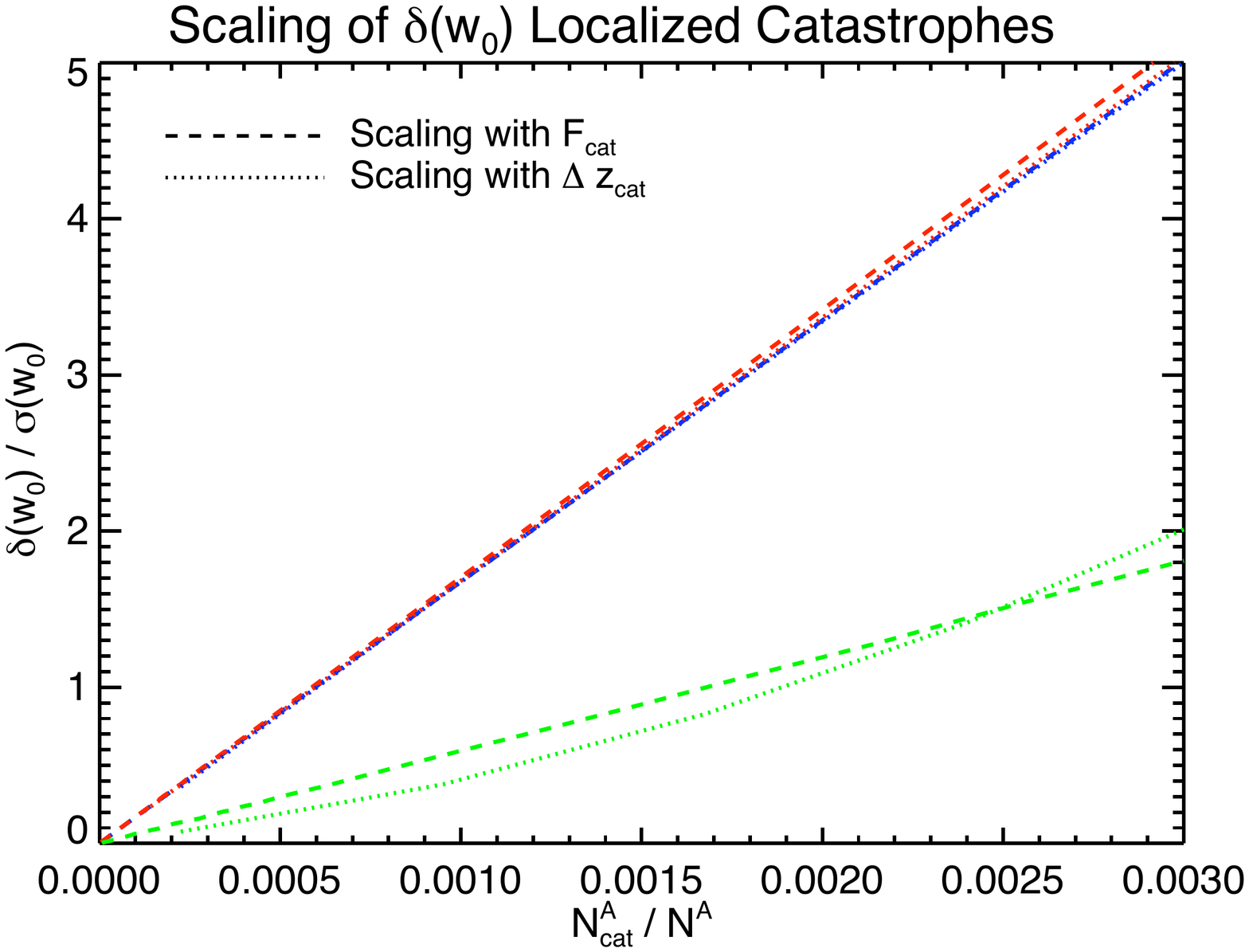}
\includegraphics[width=8.0cm]{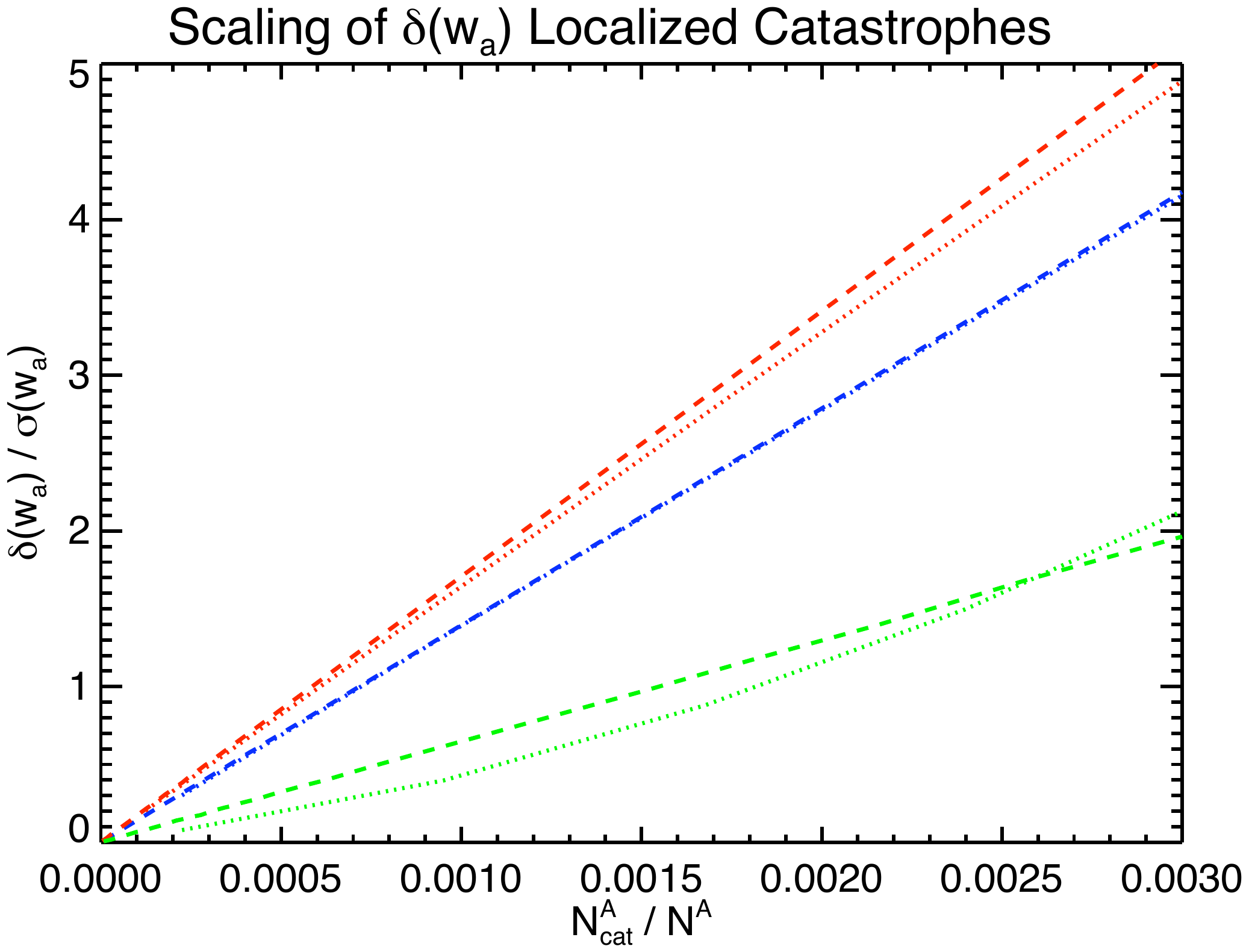}
\caption{
The scaling of systematic error in dark energy parameters with the fractional number density
 of sources whose photometric redshifts are catastrophically in error for our WIDE survey.  
On the vertical axes are the absolute value of the systematic error in $w_{0}$ (left panel) and $\wa$ (right panel) 
in units of statistical uncertainty. On the horizontal axis is $\Nacat / \Na \approx \Fcat n(\zcat)\dzcat,$
where $n(\zcat)$ is the overall redshift distribution of sources.  For the {\em dashed} curves, 
$\dzcat \equiv 0.1$ and we increase $\Nacat$ by increasing $F_{\mathrm{cat}}$.  These curves 
are all linear, as they should be.  For the {\em dotted} curves $\Fcat \equiv 0.03$ and 
we increase $\Nacat$ by increasing $\dzcat$.  These curves grow approximately as the linear, 
{\em dashed} curves.  
Three different catastrophic error localizations are color coded as 
$(\zcat,\zphcat)=(0.9,0.3)$ in red, $(0.9,2.7)$ in blue, and $(1.5,2.7)$ in green.  
The difference in intrinsic severity between these outlier populations is reflected by the slope
of the corresponding curves, with the steeper lines corresponding to the more severe systematic errors.
The agreement between {\em dashed} and {\em dotted} lines for each outliers demonstrates that the 
systematic errors induced by sufficiently well-localized catastrophes ($\dzcat \lesssim 0.3$) 
scale approximately linearly with $\Delta{}z_{\mathrm{cat}}$ over an interesting range.
}
\label{fig:scaling}
\end{figure*}

The pattern of the DES catastrophic photo-z "hot spots" differs from that of the WIDE or DEEP surveys.  
Outliers with large values of $\zphot_{\mathrm{cat}},$ that is those with a target in the fourth or 
fifth tomographic bin, are relatively more severe for DES.  
This is driven by the (assumed) comparably low redshift extent of 
imaged sources for a DES-like survey (with median redshift $z_{\mathrm{med}} = 0.7$) 
This renders a contamination that extends to high redshift more disruptive due to the small 
population of galaxies with truly high redshifts.  Though less striking, it is also evident 
in Fig.~\ref{fig:multicolor} that the DEEP survey is somewhat more sensitive than the WIDE survey
to contamination of its fourth and fifth tomographic bins.  The differences here 
are likewise driven by different survey depths and sky coverages.  A deeper, but narrower 
survey (a JDEM perhaps) is relatively more sensitive to small-scale fluctuations induced by structure 
at high-redshift, so disruptions to the higher tomographic bins are more statistically significant for DEEP 
than for WIDE. 

Finally, we return to the issue of tomographic binning with respect to the systematic errors 
in Fig.~\ref{fig:multicolor}.  We noted above that systematic errors in dark energy parameters can 
become markedly worse when the biased photometric redshifts ($\zphot_{\mathrm{cat}}$) 
distribute galaxies across the boundary of a photometric redshift bin.  The reason is 
because two sets of observables, namely the auto and cross spectra associated with the two 
target photometric redshift bins, become corrupted by the catastrophic photometric 
redshift error.  The implication is that the level of systematic error induced by a localized 
catastrophic error is quite sensitive to photometric redshift binning.  This is contrary to 
the statistical errors, which are insensitive to binning more finely than $\ntomo \approx 5$ 
over the range $0 < z < 3$ \citep{ma_etal06}.

Indeed this is the case.  The general pattern shown in Fig.~\ref{fig:multicolor} is physically 
quite sensible and is robust to binning.  However, in the case of  localized catastrophes, 
binning more finely may reduce the absolute amplitude of systematic errors if the catastrophes 
do not occur near the edge of a photometric redshift bin.  This is because smaller tomographic bins result in 
a smaller fraction of source galaxies that belong to a contaminated bin.  
This may be useful because even in the absence of significant prior indications of a localized 
catastrophe, re-analyzing the data with different photometric redshift binning schemes may 
reveal potential local catastrophes.  In the least, it should be a useful strategy to choose 
photometric redshift bins such that suspect regions of $\zphot$, where localized catastrophes may 
be anticipated, are contained in individual bins.

\subsubsection{Localized Catastrophes: Summary}
\label{sub:localsummary}

A succinct distillation of the dominant effects that determine the structure of Fig.~\ref{fig:multicolor} is as follows.  
The systematic error
induced by a localized catastrophe will be most severe when:
\ben

\item $\zcat \approx \zmed,$ which maximizes the total number of outliers; 

\item the distance between $\zcat$ and $\zphcat$ is significant;

\item and when $\zphcat$ is such that the photometric redshift bins 
contain a fractionally large contaminant (in practice, high and 
low redshift extremes).
\een
The details governing the magnitude of systematic errors generated by 
different regions of catastrophic error parameter space can be 
complicated.  In general, these details depend on the relative statistical 
weights of the affected redshift bins, as well as the characteristics of 
the survey.  

In isolation, Fig.~\ref{fig:multicolor} is useful in identifying the redshift errors that most seriously 
compromise dark energy constraints.  A shortcoming of Fig.~\ref{fig:multicolor} is that we 
have assumed catastrophic errors that occur at a fixed rate of $F_{\mathrm{cat}}=0.05$ and are active 
only over a range $\Delta{}z_{\mathrm{cat}}=0.05$.  The systematic errors induced on cosmological 
 parameters scale approximately 
with the total number of catastrophes, $N^{A}_{\mathrm{cat}}$, in Eq.~(\ref{eq:ncat}).  
In practice, scaling the systematic errors to new values of $\Delta{}z_{\mathrm{cat}}$ can be 
enacted over an interesting range of the parameter space by approximating 
$N^{A}_{\mathrm{cat}} \approx F_{\mathrm{cat}}\ n(z_{\mathrm{cat}})\ \Delta{}z_{\mathrm{cat}}$.

Figure~\ref{fig:scaling} demonstrates the validity of scaling systematic error by 
the total number of errors, $N^{A}_{\mathrm{cat}}$, for three example localized 
catastrophes.  Together, Fig.~\ref{fig:multicolor} and Fig.~\ref{fig:scaling} provide 
a blueprint for estimating the systematic error induced by a wide range of localized 
catastrophes.  One first reads off the systematic error level from Fig.~\ref{fig:multicolor}
for the grid point of interest.  
For definiteness, suppose this systematic error in either of $\wzero$ or $\wa$ is 
$\delta$.  Provided that $\dzcat$ is small, one can approximate the systematic error 
induced by a different effective value of 
$\Delta{}z_{\mathrm{cat}}$ or $F_{\mathrm{cat}}$ (call it $\delta'$) 
by scaling $\delta$ in proportion to $N^{A}_{\mathrm{cat}}$ [Eq.~(\ref{eq:ncat})], 
\begin{equation}
\label{eq:scaling}
\delta' \approx \delta \times \left({\Fcat'}/{0.05}\right) \times \left({\dzcat'}/{0.05}\right).
\end{equation}
%


In \S~\ref{sub:knowcore}, we presented results on the influence of catastrophic, uncalibrated photometric redshift errors 
on the systematic error budget for dark energy parameters $\wzero$ and $\wa$.  In that section, we assumed that the bulk 
of photometric redshifts had been well characterized by spectroscopy.  In the nomenclature of this and other papers, we 
assumed the limit in which the {\em core} of the photometric redshift distribution is calibrated so that its uncertainty does 
not contribute to the dark energy error budget.  We developed guidance on how to optimally focus photo-z calibration efforts 
and identified the most severe types of catastrophes.  In this section, we drop the assumption of arbitrarily precise calibration of 
the core populations of photometric redshifts.  Our goal is to 
assess the relative importance of calibrating the core photometric 
redshift distribution compared to eliminating catastrophic errors.  

We assume that the core photometric redshift distribution is specified by a Gaussian with redshift-dependent 
mean and dispersion.  Following \citet{ma_etal06}, we specify the unknown mean and dispersion at 31 points 
spaced evenly in redshift from $z=0$ to $z=3$ and allow for uncertainty in these parameters.   In the interest 
of simplicity, we consider a one-parameter family for the prior knowledge about the core photometric redshifts 
that may be provided by a spectroscopic calibration sample.  We do this by assuming a representative 
population of $\nspec$ galaxies with spectroscopic redshifts, distributed evenly 
in redshift from $z=0$ to $z=3,$ which can be used to calibrate the core photometric redshift distribution.  

We implement core calibration by introducing priors on the values of the dispersion and bias at 
the $i$th point in redshift.  These priors are 
\begin{eqnarray}
\label{eq:priors}
\Delta\sigma_z^{i}=\sigma_{z}^{i}\sqrt{\frac{1}{2 \nspeci}} \\
\Delta z_{\mathrm{bias}}^{i}=\frac{\sigma_{z}^{i}}{\sqrt{\nspeci}}
\end{eqnarray}
where $z_{\mathrm{bias}}^{i}$ is the bias at the $i$th point in the tabulated core distribution, $\sigma_z^{i}$ is the dispersion at 
this redshift, and $\nspeci$ is the number of spectroscopic galaxies in each of the 31 bins of width $\delta z=0.1$ used to calibrate 
the core photo-z redshift distribution. 
 This prior model is certainly simplistic.  For example, in our analysis we have chosen for the sake of simplicity to set all of the $\nspeci$ equal to each other,
 so that our implementation assumes that calibrating spectra 
are sampled equally in redshift,
  whereas in reality we will have much looser constraints on sources at high redshift than 
those at low redshift.  Moreover, both core calibration {\em and} the ability to identify catastrophic outliers improve with 
larger spectroscopic samples.  However, we consider these issues independently in the interest of completeness because 
the details of how a realistic calibration program may proceed remain uncertain. 

\subsection{Catastrophic Redshift Errors with Core Uncertainty}
\label{sub:unccore}

\begin{figure*}[t!]
\centering
\includegraphics[width=8.0cm]{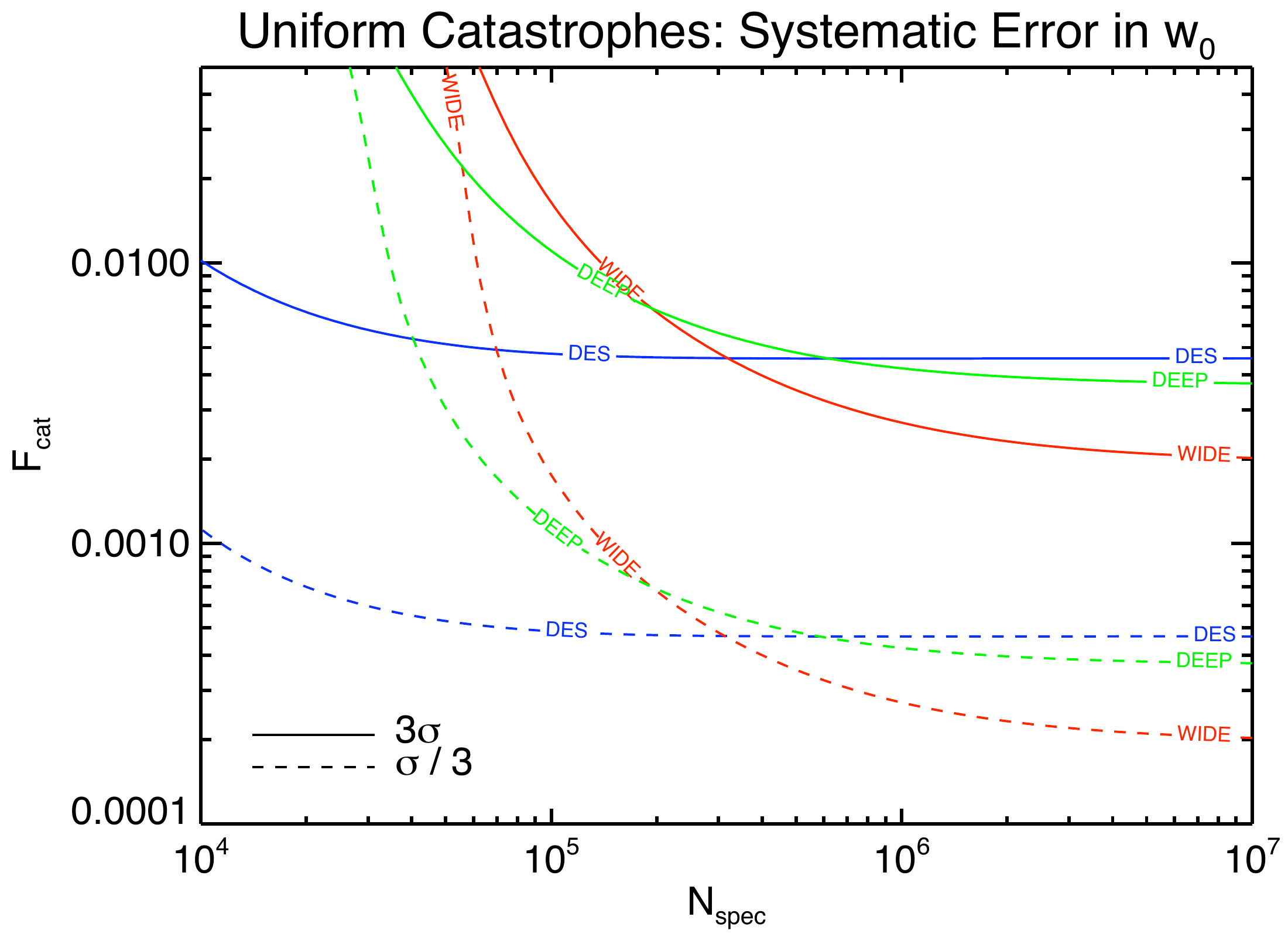}
\includegraphics[width=8.0cm]{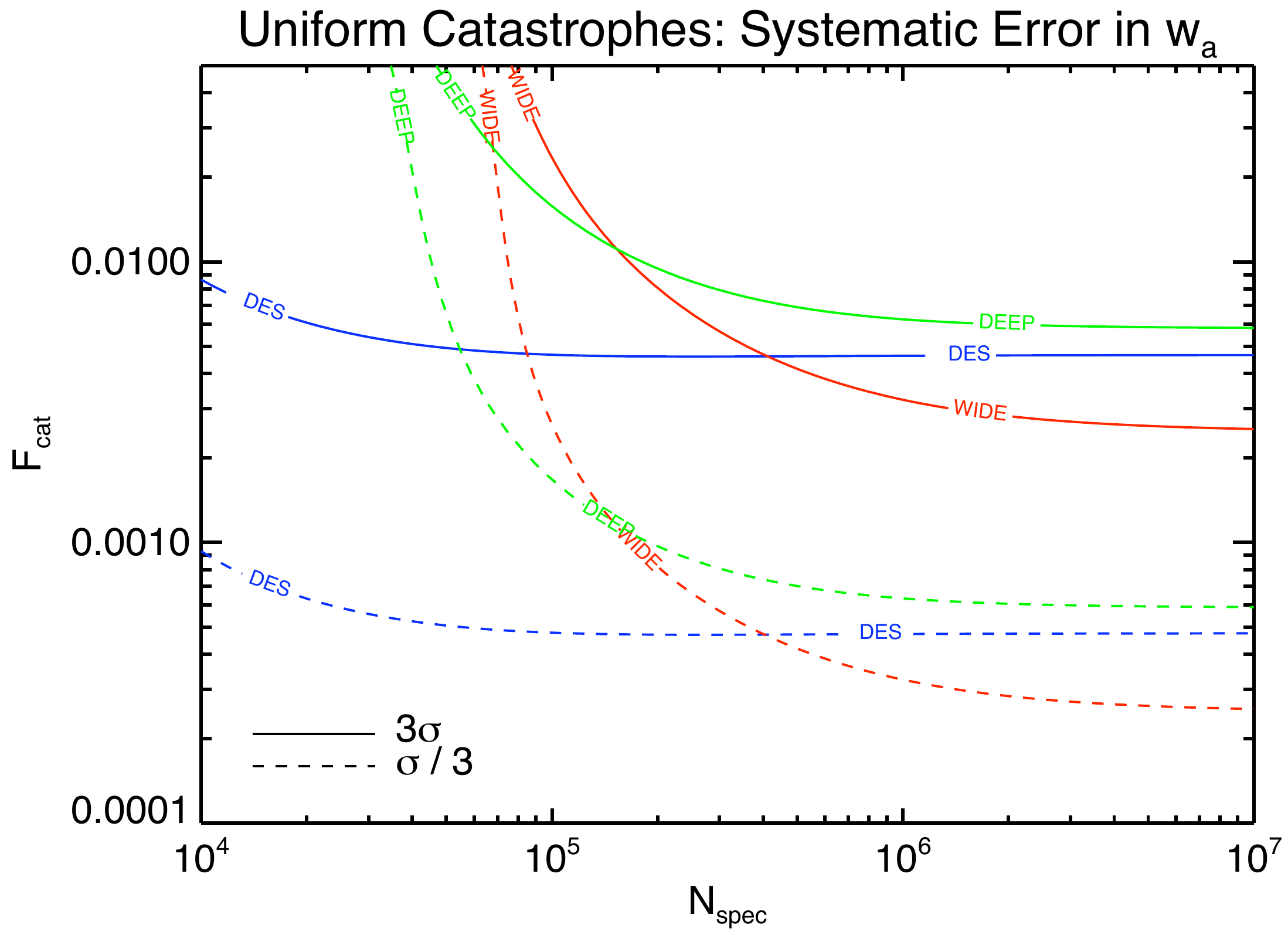}
\includegraphics[width=8.0cm]{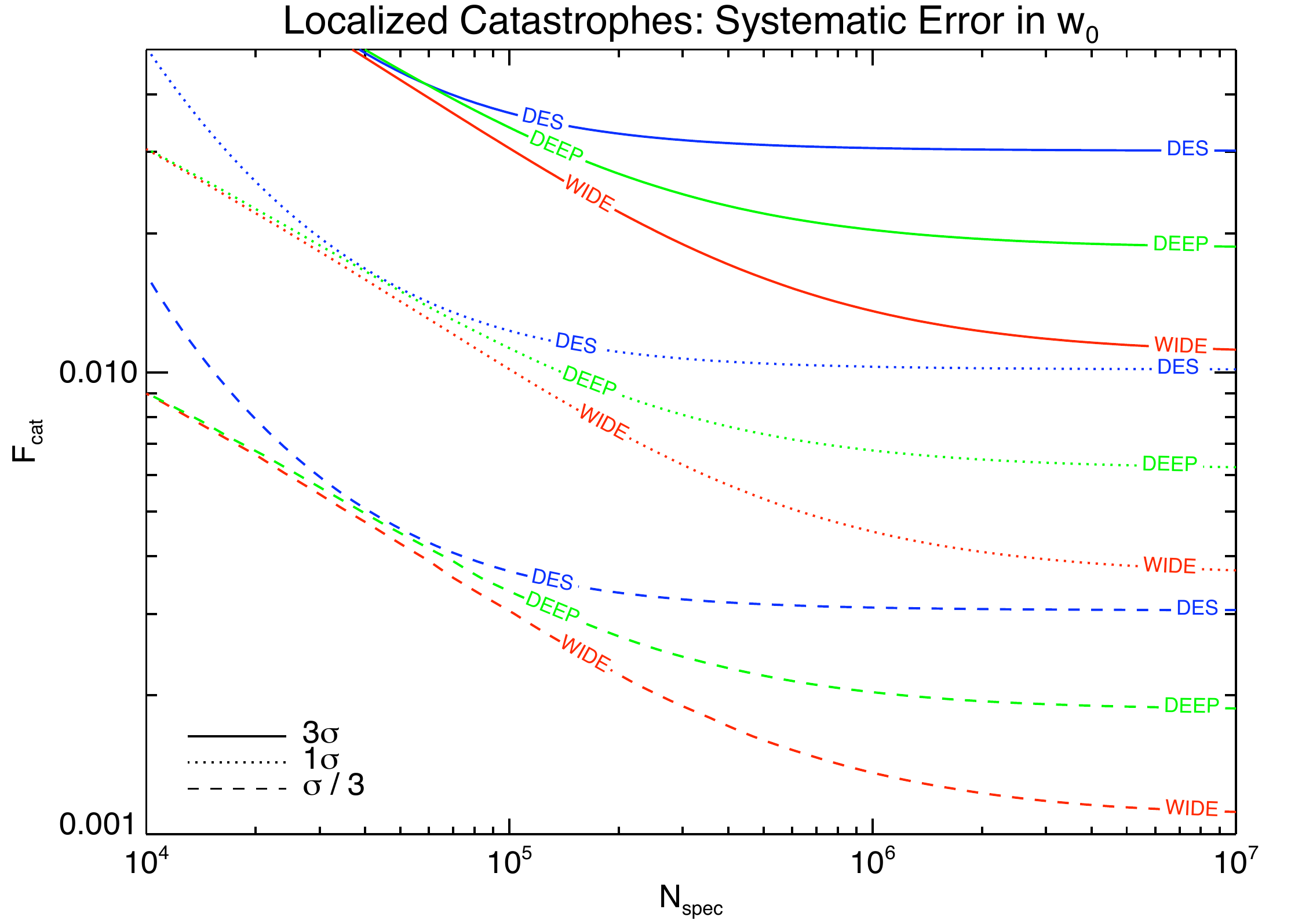}
\includegraphics[width=8.0cm]{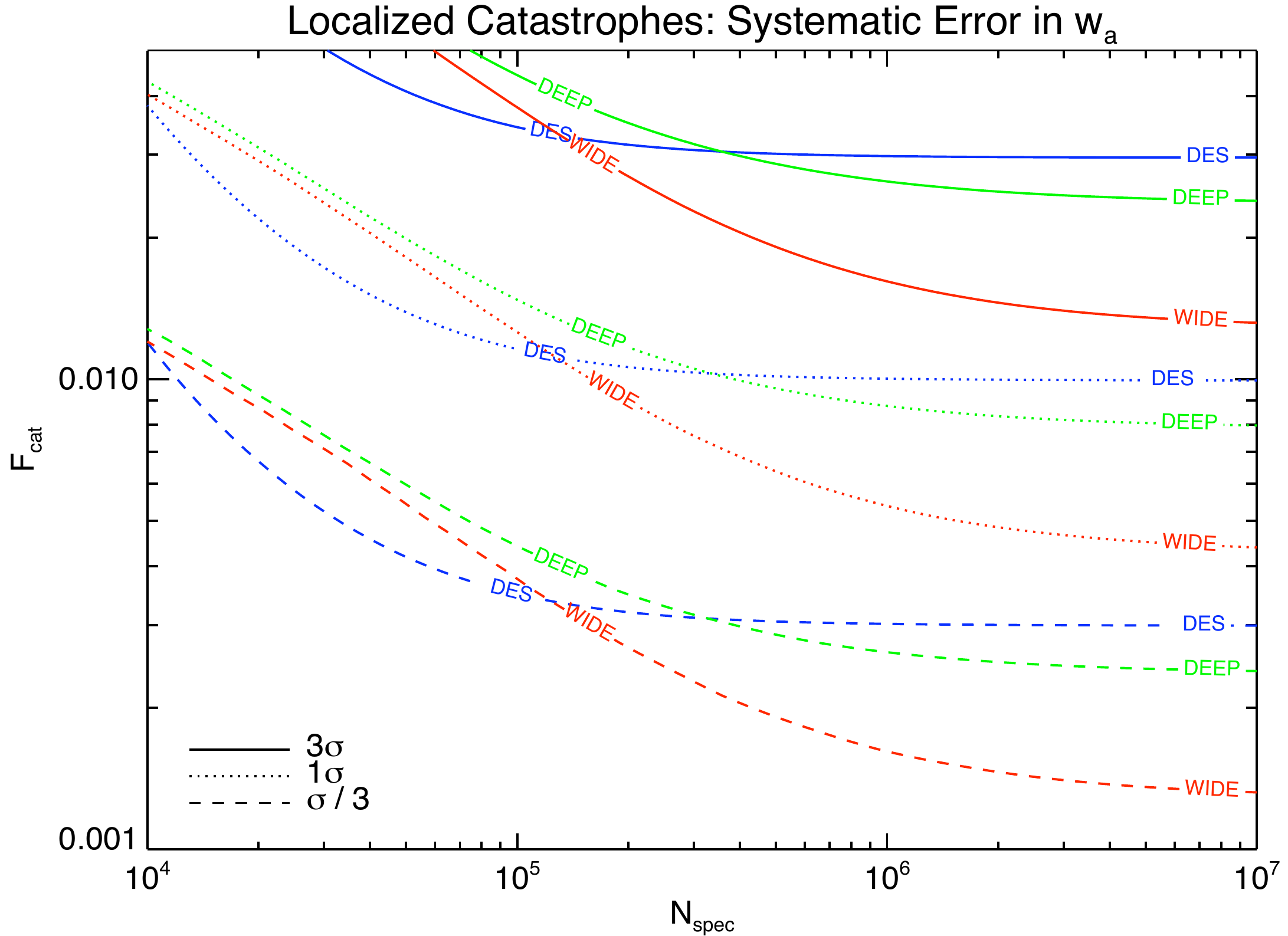}
\caption{
Contours of constant $w_{0}$ and $w_{a}$  bias from the worst case catastrophe in units of the statistical uncertainty of the survey.  
Systematic errors in $w_{0}$ appear in the {\em left} panels and $w_{a}$ in the {\em right} panels.  Results for the worst case uniform 
catastrophe appear in the {\em upper} panels, and were generated with $\Delta{}z_{cat}=1.5$ and $z_{cat}=1.5.$  
Contours of systematic error produced by localized catastrophes appear in the {\em bottom} panels.
Each of the localized contours have been calculated with 
$\Delta{}z_{cat}=0.1$ and $\sigma_{cat}=0.03$.  For DEEP and WIDE $z_{cat}=1.15.$ and $z_{cat}^{ph}=0.15,$ and for 
DES $z_{cat}=0.85,$ and $z_{cat}^{ph}=0.15,$ in accordance with the results illustrated in Fig.~\ref{fig:multicolor}.}
\label{fig:worstcase}
\end{figure*}

Figure \ref{fig:worstcase} is a contour plot depicting the systematic errors in $\wzero$ and $\wa$ induced by the worst-case-scenario 
catastrophes determined in \S~\ref{sub:knowcore}.  The prior core knowledge is specified by $\nspec$, which runs along the 
horizontal axis.  The error rate, $F_{\mathrm{cat}},$ runs along the vertical axis.  
For uniform catastrophes the worst case outliers span the true redshift range of 
the survey.  For localized catastrophes the most sinister outliers lie at the points of 
maximum systematic error in Fig.~\ref{fig:multicolor}.  
The dashed (solid) curves are lines of constant systematic error at a level of one-third (three times) 
the statistical error on each parameter.
Clearly then, systematic errors are dominant above the solid curves and become unimportant well below the 
dashed curves.  
In the bottom panels we have included dotted curves to emphasize the region of parameter space
where systematic errors are equal to statistical errors.  These $1\sigma$ contours are omitted in the upper panels
to avoid clutter, but the linear dependence of the induced systematic error on the catastrophic error rate 
ensures that the $1\sigma$ can be estimated by scaling the $3\sigma$ or $\sigma/3$ contours by a 
factor of three.

Several aspects of Fig.~\ref{fig:worstcase} are worthy of note.  The contours all become very flat at large $\nspec$.  
This is the limit in which the core photo-z distribution is calibrated sufficiently well that it no longer contributes to the 
error budget of $\wzero$ and $\wa$ \citep[e.g.,][]{ma_etal06,ma_bernstein08}.  This corresponds to the limit of 
perfect knowledge of the core photo-z distribution, and accordingly, the systematic errors asymptote to those quoted in 
\S~\ref{sub:knowcore} at large $\nspec$.

For a fixed level of systematic error, experiments generally become less tolerant of catastrophic outliers as $\nspec$ increases.  
This behavior is reflected in the negative slope at the low $\nspec$-end of the contours of constant systematic error in 
Fig.~\ref{fig:worstcase}.  This is an explicit manifestation of the competition between calibration of the "core" population of 
photometric redshifts and the ability to diagnose and eliminate a sub-dominant, poorly-understood "catastrophic" outlier population.  
The reason for this is simply that {\em systematic} errors must be better understood for samples with smaller {\em statistical} uncertainty.  
If the statistical errors in the measurement are intrinsically large, as they would be in the 
limit of poorly-calibrated photo-z's for the majority of the imaging sample, then high-rates of catastrophic outliers are 
tolerable because the systematic they contribute is not large compared to the statistical error induced by 
a poorly constrained core distribution.  

When the core distribution is very well calibrated, most obvious at $\nspec \gtrsim 10^5$ for uniform catastrophes in DES, the 
contours of constant systematic error transition to slightly positive slope.  This occurs when the 
core distribution has been sufficiently well calibrated that degeneracies between the photometric 
redshift parameters of the core distribution and cosmological parameters are no longer significant.  
Calibrating beyond the level required to break degeneracies between cosmology and the core 
photo-z parameters results in a slight reduction in systematic errors on cosmological parameters.  
This is a specific manifestation of the general result that improving priors can only lead to a net 
reduction in the systematic errors of inferred parameters, a result discussed in considerable 
detail in \citet{bernstein_huterer09}.  Clearly, the reduction in systematic error at very large 
$\nspec$ is not significant in the cases of interest here.

To illustrate the competition between core calibration and the removal of outliers, consider some explicit examples.  
In the case of the uniform catastrophe, our Deep (Wide) survey can tolerate 
catastrophic errors at a rate $F_{\mathrm{cat}} > 1\%$ if the core calibration is worse than the statistical equivalent of 
$\nspec \lesssim 3 \times 10^4$ ($\nspec \lesssim 6 \times 10^4$).   For both surveys, even the worst-case, localized 
catastrophes can occur at a rate of $F_{\mathrm{cat}} > 1\%$ if $\nspec \lesssim 10^4$.  Of course, 
the worst-case localized systematic errors are more subtle to interpret, as we have assumed they are only actively 
affecting galaxies over a range of true redshifts with width $\Delta{}z_{\mathrm{cat}} = 0.1$; 
however, the magnitude of the induced systematic errors produced by localized catastrophes active over 
different redshift ranges scales in proportion to 
$N^{A}_{\mathrm{cat}} \sim n(z_{\mathrm{cat}}) \Delta{}z_{\mathrm{cat}} F_{\mathrm{cat}}$, as illustrated in Fig.~\ref{fig:scaling}.  
Detailed results are complex, but two simple conclusions are clear:
\begin{enumerate}
\item  Limiting uniform catastrophic error rates to $F_{\mathrm{cat}} \lesssim 4 \times 10^{-4}$  $(F_{\mathrm{cat}} \lesssim 2 \times 10^{-4})$ 
for DES and DEEP (WIDE) will render them unimportant.  
\item  Limiting individual localized catastrophic error rates to 
$F_{\mathrm{cat}} (\Delta{}z_{\mathrm{cat}}/0.1) \lesssim 10^{-3}$ will 
render them unimportant for each experiment.
\end{enumerate}
 
In practice, some amount of uncertainty in the calibration of the core distribution is inevitable, so error 
rates higher by a factor of a few may be tolerable, but in detail this will depend upon 
the nature of the error and the properties of the core sample of well-calibrated photometric 
redshifts.  Fig.~\ref{fig:multicolor} and Fig.~\ref{fig:worstcase} contain the information necessary 
to diagnose the systematic error for a variety of idealized, but interesting cases.

\subsection{Mitigating Systematic Errors by Sacrificing Statistics}
\label{sub:mitigation}

In \S~\ref{sub:knowcore} and \S~\ref{sub:unccore}, we estimated the systematic errors that could be induced 
by two broad families of catastrophic photometric redshift error, remaining relatively agnostic about the source 
of the error.  We found generally that error rates must be kept to levels below $F_{\mathrm{cat}} \sim 10^{-3}$, or 
one of a thousand imaged galaxies with large, uncalibrated redshift errors in order for systematic errors 
not to contribute to the dark energy error budget (though specific tolerances depend upon several details).  
This will be a relatively challenging goal for a photometric redshift calibration 
program to attain.  DES, JDEM, EUCLID and LSST will all require calibration of very 
faint galaxies, where precise photo-z's are difficult to obtain.  Moreover, the types of galaxies 
imaged, and for which spectra may be available, varies as a function of redshift, 
so some understanding of the details of galaxy evolution 
will be needed in order to achieve calibration goals.  

It is natural to explore simple methods to sacrifice some of the 
statistical power of imaging surveys in order to mitigate larger systematic errors.  One of the simplest 
techniques we can employ to limit the effect of catastrophic outliers is to place cuts on the range of 
photometric redshifts utilized to infer cosmological parameters (\citealt{bernstein_huterer09} have explored such 
cuts for a particular model of photo-z outliers).\footnote{\citet{nishizawa_etal10} also study the ability to employ photometric
redshift cuts to mitigate the effects of catastrophic outliers, 
which became available on the Arxiv while we were submitting this manuscript for publication.}  The most damaging 
catastrophic errors are those that take galaxies near the median redshift of the survey and scatter 
them to significantly lower or higher redshifts, so it is sensible to explore the losses in statistical 
power incurred by excising galaxies at the low- and high-redshift ends of surveys.

\begin{figure*}[t]
\centering
\includegraphics[width=8.5cm]{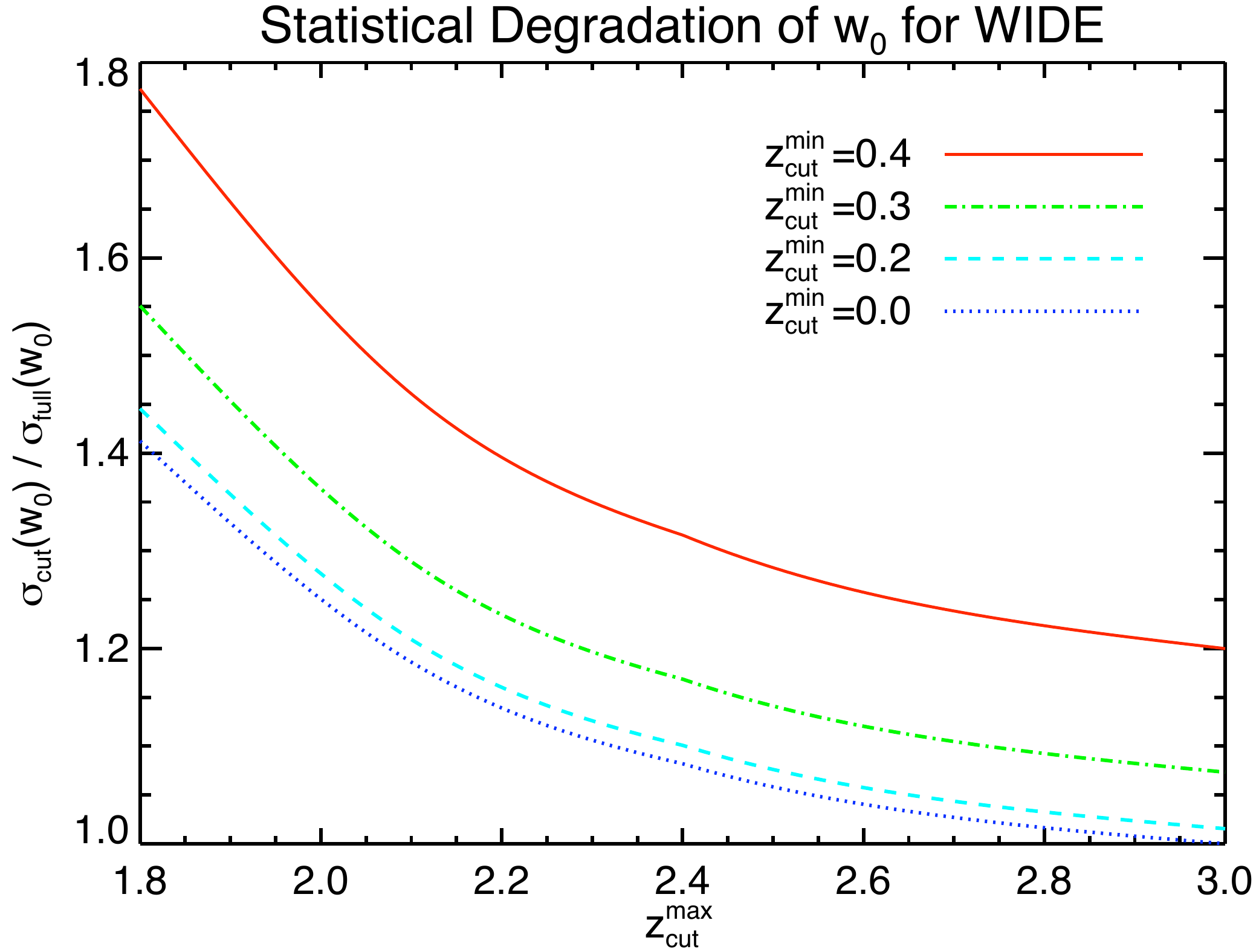}
\includegraphics[width=8.5cm]{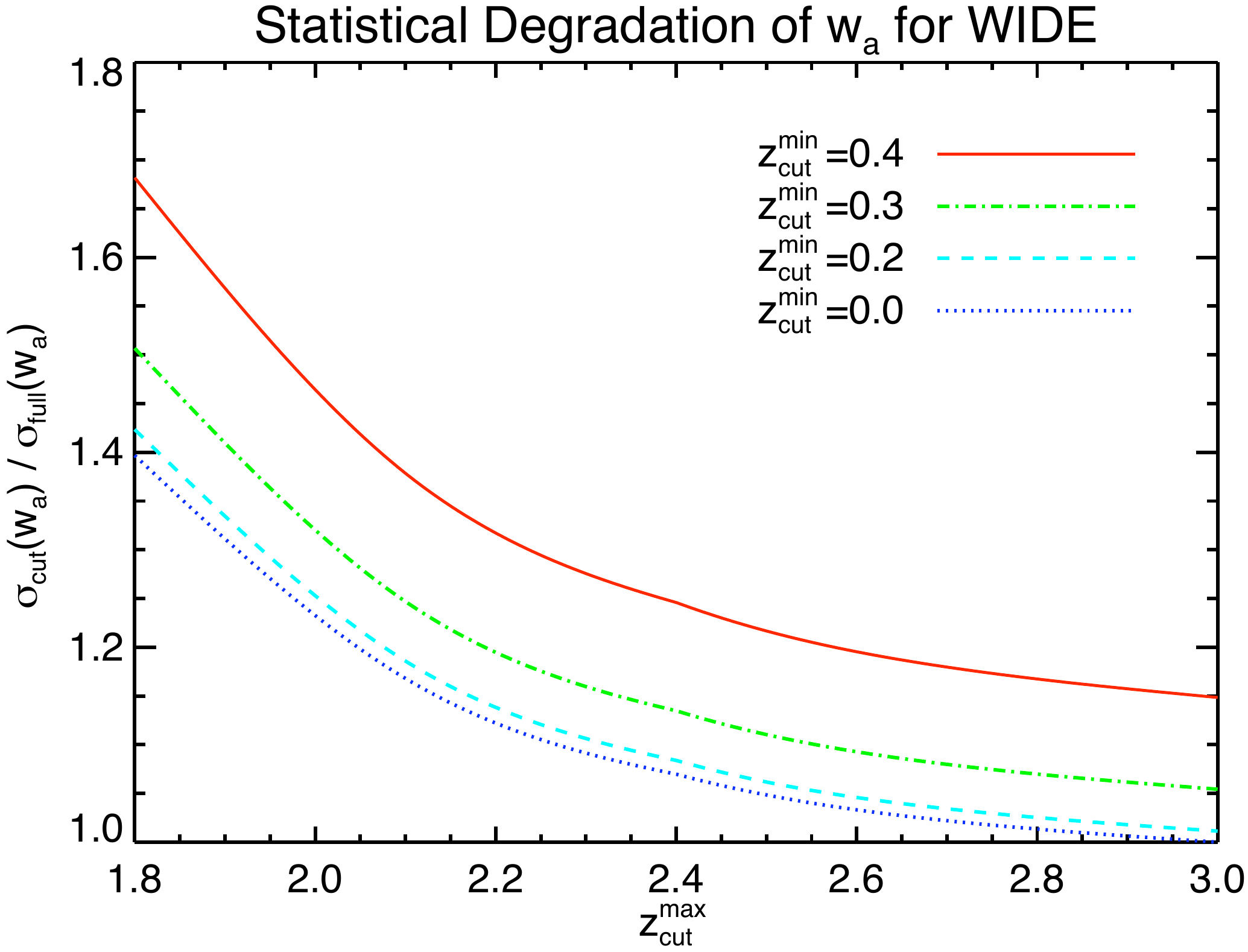}
\includegraphics[width=8.5cm]{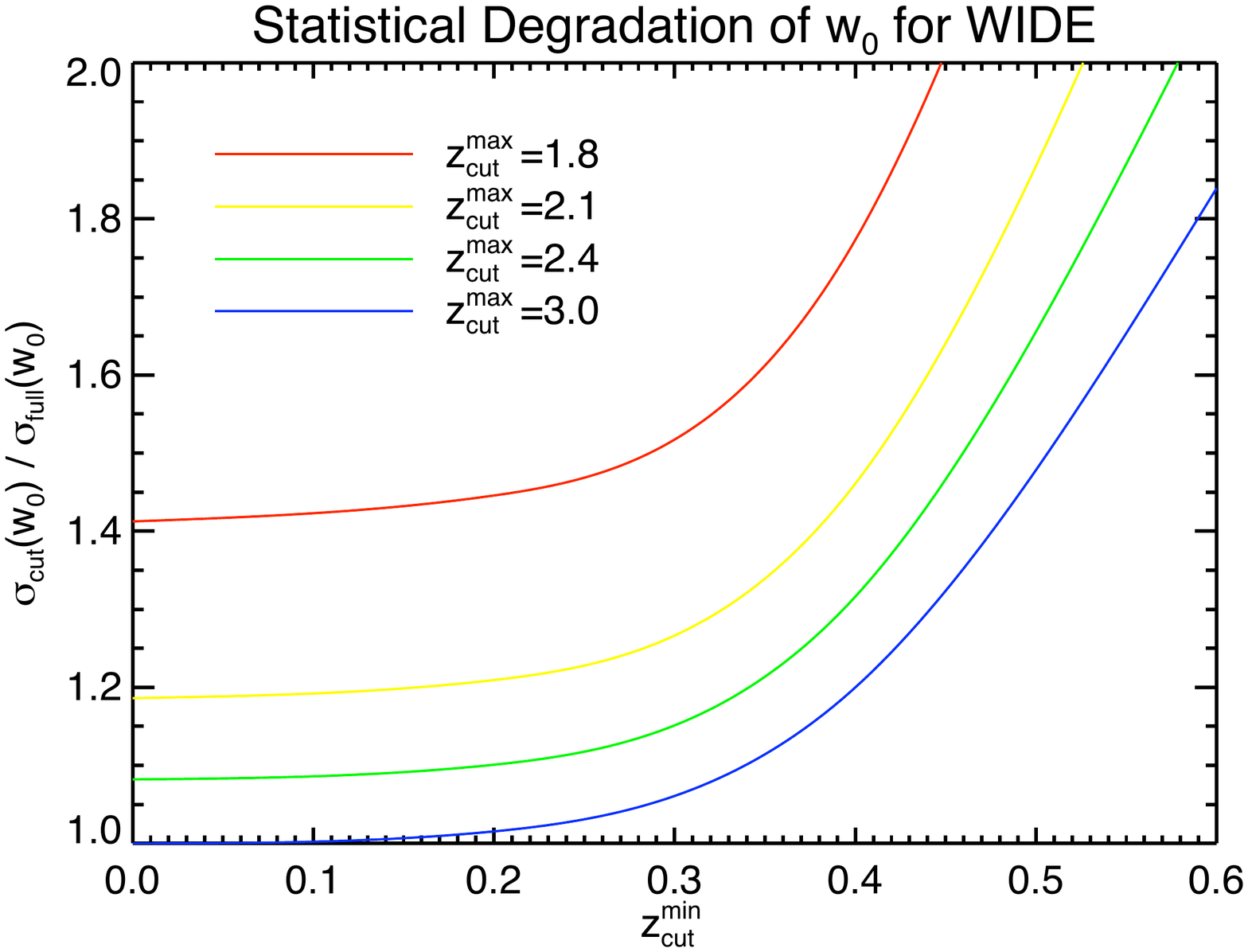}
\includegraphics[width=8.5cm]{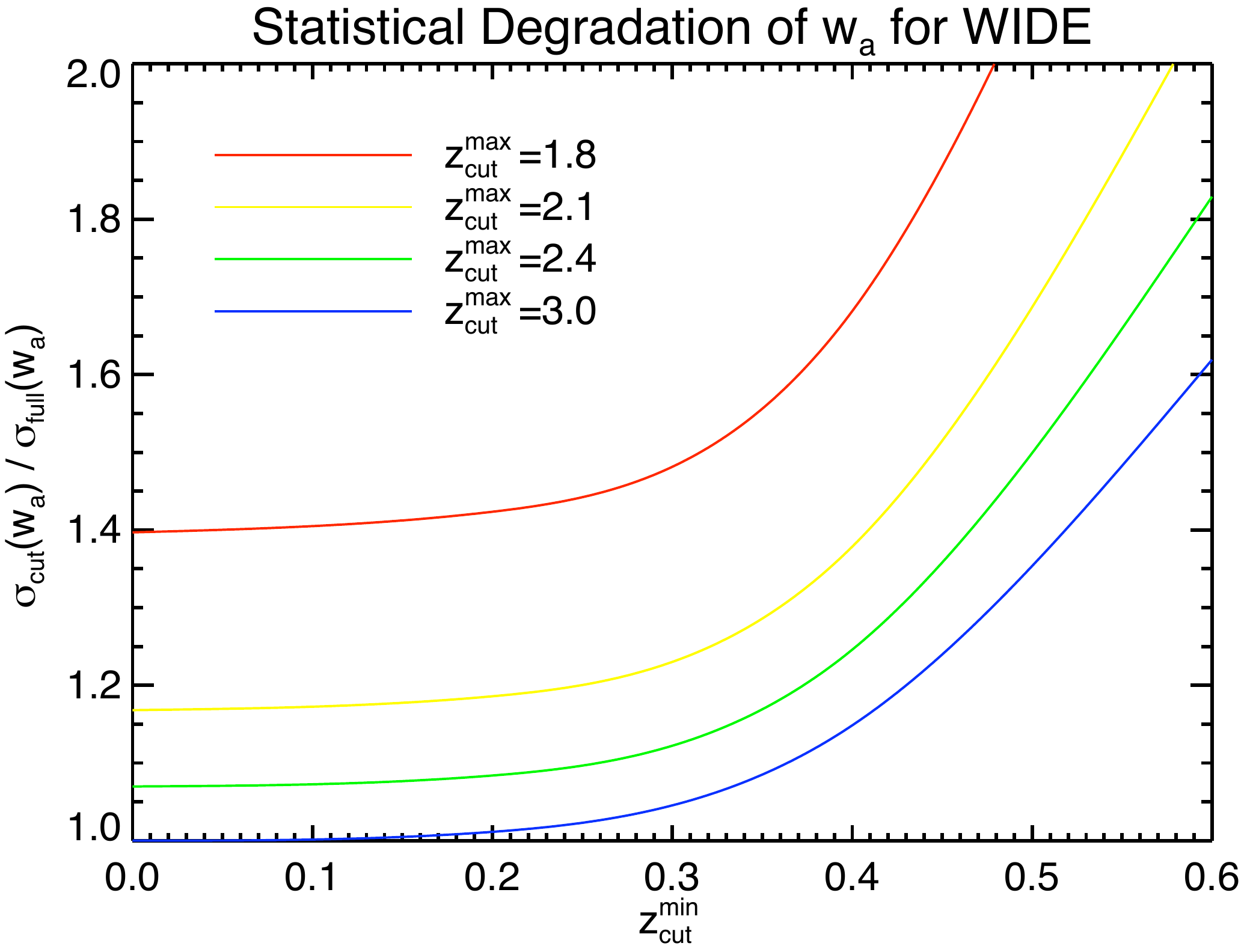}
\caption{
The statistical cost of excising low- and high-redshift shear information on constraints of 
$w_{0}$ ({\em left panels}) and  $\wa$ ({\em right panels}) for our WIDE survey. 
In the {\em top} row, the value of the maximum photometric redshift of the survey appears 
along the horizontal axis while the different lines show different choices of the minimum 
photometric redshift as indicated.  Along the vertical axis is the fractional increase in dark energy 
parameter constraints relative to the constraints provided by a survey with our standard tomography.  
In the {\em bottom} row, the value of the minimum photometric redshift of the survey runs 
along the horizontal axis while the different lines show different choices of the maximum 
photometric redshift as indicated.
}
\label{fig:squeeze}
\end{figure*}

We demonstrate the utility of photometric redshift excision in this section by exploring a class 
of simple excision algorithms.  In particular, we cut out all galaxies with {\em photometric} 
redshifts greater than some value, $z_{\mathrm{max}}^{\mathrm{cut}}$, and smaller 
than some value $z_{\mathrm{min}}^{\mathrm{cut}}$.  
Figure~\ref{fig:squeeze} shows the statistical errors on $\wzero$ and $\wa$ 
as a function of $z_{\mathrm{max}}^{\mathrm{cut}}$ and 
$z_{\mathrm{min}}^{\mathrm{cut}}$ for our Wide survey, whose characteristics
are similar to those expected from an LSST- or Euclid-like survey.  
The relative costs depend mildly upon survey parameters.

Excising galaxies with photometric redshifts lower than 
$\zphot \sim 0.3$ results in only a $\sim 7\%$ increase in the statistical 
errors on dark energy parameters.  Likewise, excising galaxies with 
$\zphot \gtrsim 2.4$ results in only a $\sim 10\%$ degradation in $\wzero$ 
and $\wa$ constraints.  Excising both of these regions of 
photometric redshift leads to a reduction in constraining power of $\lesssim 20\%$.  
Fig.~\ref{fig:squeeze} is a valuable itemization of the statistical losses incurred by 
redshift cuts and indicates that excising low- and high-redshift 
portions of the imaging surveys may be an effective method to mitigate the influence of 
catastrophic photometric redshift errors at little cost in statistical error.

While Fig.~\ref{fig:squeeze} quantifies the cost of excising regions of photometric redshift, 
the parametric complexity of catastrophic photo-z errors makes specific statements 
about the benefit of such cuts more difficult.  In the case of a {\em localized} catastrophe that places 
galaxies erroneously in the excised high- or low-redshift ends of the survey, 
the induced bias can be nearly completely removed at the cost of the statistical 
degradation in Fig.~\ref{fig:squeeze}.  We have begun a preliminary 
study of the benefits of redshift excision, including the case of {\em uniform} 
catastrophes.  In the case of our WIDE survey, excision can considerably 
reduce systematic errors induced even by the {\em uniform} catastrophe when 
the core is not well-calibrate ($\nspec \lesssim 10^5$), but this strategy is 
only of marginal value in the limit of a well-calibrated core.  We limit the 
present discussion to the itemization in Fig.~\ref{fig:squeeze} and relegate 
further study of redshift cuts and possible self-calibration of specific types 
of catastrophic error to a follow-up study.

\section{Conclusions and Discussion}
\label{section:discussion}

We have studied the potential systematic errors that may be induced in dark energy 
parameters inferred from forthcoming weak lensing surveys as a result of a population 
of source galaxies with photometric redshifts that deviate significantly from their true 
redshifts.  We used a particular operational definition of catastrophic photo-z errors 
that is subtly distinct from the use of this term in some of the existing literature.  
Throughout this work, the term {\em catastrophic photometric redshift error} refers to cases in which 
photo-z estimates differ significantly from true redshifts, the nature of the error has not been identified 
or calibrated with an accompanying spectroscopic data set,  {\em and}  the outlier population 
has not been removed reliably from the imaging data prior to the construction of 
shear correlation statistics.  One way to interpret our results 
is as requirements for spectroscopic calibration of outliers and the completeness 
with which outlier galaxies must be culled from the data set in order 
to render systematic errors in dark energy parameters small.

In order to provide relatively general guidelines on the fidelity with which 
outlier photo-z's must be understood, we have taken an agnostic position 
on the nature of what types of catastrophic photometric redshift outliers may 
be realized in forthcoming imaging data.   This eliminates the need to anticipate 
what types of photo-z errors may occur at very small fractional rates in order to 
assess their general influence on dark energy parameters.  
To be sure, there are reasonable guesses that can be made 
regarding the nature of photometric redshift errors and 
many algorithms exist that estimate redshifts from photometric 
data and refine estimates based upon comparisons with large, spectroscopic 
data sets \citep[e.g.,][]{bolzonella_etal00,collister_lahav04,oyaizu_etal07,feldmann_etal06,
brammer_etal08,margoniner_wittman08,cunha_etal09,ilbert_etal09,coupon_etal09}.  
However, we have not adopted any particular template for photometric 
redshift outliers.  
Instead, we have studied two extreme limiting cases of catastrophic photometric redshift error.

In the first class of photometric redshift error, which we dubbed the {\em uniform} 
catastrophe, photometric redshifts are poorly constrained and scattered over a 
broad range \citep[see, e.g.,][for examples of such features]{ilbert_etal09,coupon_etal09}.  
Photo-z errors resembling our uniform type must be well controlled.  If such errors occur even 
for a relatively small fraction of galaxies near the median redshift of a given survey, the systematic 
errors induced on dark energy parameter estimators will be significant.  Roughly speaking, 
we find that the error rate per galaxy must be maintained at 
$F_{\mathrm{cat}} \lesssim \mathrm{a}\ \mathrm{few} \times 10^{-4}$.  
However, the uniform catastrophic error is a relatively simple variety so that self-calibration 
may well be feasible.  One could resign oneself to the fact that such an error will occur and add the error 
rate $F_{\mathrm{cat}}$ (and perhaps other parameters such as $\Delta{}z_{\mathrm{cat}}$) to the set of 
nuisance parameters to be marginalized over.  This self-calibration could eliminate the systematic error, but 
will broaden statistical errors.  We explore self-calibration of particular catastrophic photo-z errors 
in a forthcoming paper.

The second class of errors, which we refer to as {\em localized} catastrophes, takes source galaxies with 
particular true redshifts and assigns them photometric redshifts with a large bias but small 
scatter.  Localized catastrophes have a broader range of possibilities and are more difficult to deal with.  
Fig.~\ref{fig:multicolor}, Fig.~\ref{fig:worstcase}, and Eq.~(\ref{fig:scaling}) constitute a blueprint 
for estimating the severity of a broad range of possible localized photometric redshift catastrophes.  
Quite generally, we find that the systematic errors they induce are sensitive to the scheme 
used to bin the source galaxies in photometric redshift.  This suggests that an iterative scheme 
of re-binning may be an effective strategy for identifying and mitigating the influence of 
localized catastrophic photo-z errors.

In \S~\ref{sub:mitigation} we studied a simple strategy to limit the systematics induced by 
catastrophic photo-z outliers.  First, we showed that the statistical leverage of the highest redshift ($z \gtrsim 2.4$) 
and lowest redshift ($z \lesssim 0.3$) source galaxies on dark energy constraints is minimal.  
Eliminating all such galaxies from consideration in inferring dark energy parameters results in only a 
small increase in the statistical errors of dark energy equation of state constraints, but may eliminate 
some of the most severe systematic errors induced by localized catastrophic photo-z outliers.  
This implies that well-designed cuts on $\zphot$ will likely be a powerful and general means to mitigate 
systematics associated with photo-z determination at a relatively small cost in statistical error.  

The published work that is most closely related to the present work is \citet{bernstein_huterer09}.  
Our work is an extension and generalization of their study.  Overall, we reach the same 
broad conclusions where the two studies are commensurable.  In particular, 
we find that catastrophic errors of the localized variety must be controlled 
such that the rate of errors per galaxy is $F_{\mathrm{cat}} \lesssim 10^{-3}$ if 
they are to induce tolerable systematic errors on dark energy parameters.  

Our work differs from and complements \citet{bernstein_huterer09} in several important ways.  
First, we have relaxed the assumption that 
the true redshift distribution of the outlier population perfectly traces that of the core population within 
individual Source photometric redshift bins (see Eq. \ref{eq:pgausscat}).  Our treatment of photometric 
redshift errors is independent of the photometric redshift binning (as such errors would be in practice), 
while the approach of \citet{bernstein_huterer09} is limited to cases 
in which photometric errors both trace the galaxy distributions within the Source Bin and span the redshift
range of the Source Bin.  
While contamination of the Target redshift bin is typically the larger source of 
induced systematic error, our generalization illustrates that the effects of modifications to the 
Source Bin are non-negligible and in some cases these offsets contribute significantly to 
the systematic errors on dark energy parameters.  
Second, we have studied catastrophic errors in cases 
where the {\em core} photometric redshift distribution is not perfectly calibrated.  Accounting for 
uncertainty in the core distribution turns out to be quite important: for a fixed catastrophe the 
magnitude of the induced systematic errors can vary by several orders of magnitude over a 
reasonable range of priors on the core distribution.  Third, we have explored cases of correlated shifts in photo-z errors that 
span multiple tomographic redshift bins (which will occur in practice), the extreme example being 
the {\em uniform} error.  

We conclude our discussion section by referring to interesting, 
tangential results given in the appendix.  In the appendix, we 
discuss the effect of different models of the nonlinear evolution 
of cosmological density perturbations on photometric 
redshift calibration requirements.  Weak lensing measurements take significant advantage 
of measurements on nonlinear scales in order to constrain cosmology.  Previous work on 
the calibration of photometric redshifts has utilized the \citet{peacock_dodds96} 
formula \cite[e.g.][]{ma_etal06,ma_bernstein08}; however, 
we find that using the more recent and more accurate fit of \citet{smith_etal03} significantly 
reduces the need for independent calibration of photometric redshifts.  We have used the 
\citet{smith_etal03} formula in the main body of this paper.  We refer 
the reader to the Appendices for further details.

\section{Summary}
\label{section:summary}

We have adopted a simple, agnostic approach to estimate the levels at which uncalibrated 
photometric redshift outliers must be controlled to maximize the dark energy constraints from 
the weak lensing components of forthcoming imaging surveys such as DES, LSST, EUCLID, and JDEM.  
We present results for three fiducial imaging surveys: a relatively near-term DES-like survey; a future survey 
with a high surface density of galaxies but a relatively small fractional sky coverage (DEEP); and a 
future survey with half-sky coverage and a lower galaxy surface density (WIDE).  
We considered two extreme cases of large, uncalibrated errors.  In the case of a 
{\em uniform} photo-z catastrophe, we considered galaxies erroneously assigned photometric redshifts 
that are unrelated to their true redshifts.  In the case of a {\em localized} photo-z catastrophe, 
we considered the erroneous placement of a small fraction of galaxies in some range of 
true redshifts at significantly different {\em photometric} redshifts.  
To be specific, we assigned galaxies in some range of true redshifts of width 
$\Delta z$ centered on a true redshift $z_{\mathrm{cat}}$ to photometric redshifts near 
$\zphot_{\mathrm{cat}}$ that differ significantly from $z_{\mathrm{cat}}$.  
For each type of error and survey, we assessed the severity of the 
systematic errors on dark energy parameters that would be 
induced by catastrophic photometric redshift errors.  
Our primary results are as follows.

\begin{enumerate}

\item A photometric redshift error of the {\em uniform} variety that is 
relevant for galaxies near the median redshift of the imaging survey, 
must be limited to a fraction of galaxies $F_{\mathrm{cat}} \lesssim 5 \times 10^{-4}$ 
for DES or DEEP and $F_{\mathrm{cat}} \lesssim 2 \times 10^{-4}$ for WIDE, 
in order to induce systematic errors that are small compared to 
the statistical errors on $\wzero$ and $\wa$.  

\item {\em Localized} catastrophic errors are most severe when they take some  
fraction of galaxies with true redshifts near the median survey redshift and 
assign them significantly higher or lower photo-z's.  For DES, assignments 
to higher photo-z's are more severe than assignments to lower 
photo-z's while the opposite is true for WIDE and DEEP.  However, the systematic 
errors induced by these two extremes differ by less than a factor of two in all cases.  

\item Limiting the fraction of galaxies exhibiting {\em localized} catastrophes at all 
redshifts to $ F_{\mathrm{cat}} \lesssim 3 \times 10^{-3}$ for DES or 
$F_{\mathrm{cat}} \lesssim 10^{-3}$ for WIDE or DEEP will render them unimportant.  
For localized catastrophes that occur over a range of true redshifts of width 
$\Delta{}z_{\mathrm{cat}}$ near the median survey redshift, 
the fractional error rate must be controlled such that 
$F_{\mathrm{cat}} (\Delta{}z_{\mathrm{cat}}/0.1) \lesssim  1-3 \times 10^{-3}$.  

\item Imperfect knowledge of the photo-z distribution for the {\em core} sample of galaxies 
loosens these requirements for uncalibrated catastrophic outlier control as depicted in 
Fig.~\ref{fig:worstcase}.  Roughly speaking, core calibration with spectroscopic 
samples smaller than the statistical equivalent of $\nspec \lesssim 10^5$ leads to 
significantly reduced catastrophic error control requirements.  
Of course, in practice catastrophic error control and core calibration 
will both improve as $\nspec$ increases.

\item The statistical leverage of the highest redshift ($z \gtrsim 2.4$) 
and lowest redshift ($z \lesssim 0.3$) source galaxies on dark energy constraints is 
small.  Eliminating all such galaxies from consideration in inferring dark energy parameters 
results in a $\lesssim 20\%$ increase in the statistical errors on dark energy, 
but may eliminate the most severe systematic errors induced by localized catastrophic 
photo-z outliers.  

\item  In the appendix, we show that dark energy parameter forecasts 
that include photometric redshift uncertainty vary significantly depending upon the 
treatment of the nonlinearity in the matter power spectrum.  In particular, using the 
\citet{smith_etal03} fitting form (as we do in the main text) leads to weaker 
photo-z calibration requirements than does the \citet{peacock_dodds96} 
formula upon which the results of \citet{ma_etal06} are based.  The \citet{smith_etal03} 
formula has been shown to be more accurate than \citet{peacock_dodds96} suggesting 
that degradation due to photo-z uncertainty may be less than \citet{ma_etal06} forecast.  
Only a rigorous numerical study can determine this definitively.  

\end{enumerate}

This level of photometric redshift outlier control is challenging in comparison to 
the yields of contemporary methods and data.  
Existing spectroscopic samples are not representative of the galaxy populations
that will be utilized to constrain dark energy in forthcoming imaging surveys.
For example, the Deep Extragalactic Evolutionary Probe (DEEP2) 
has a $70\%$ success rate for obtaining spectroscopic redshifts \citep{cooper_etal06}, 
where star-forming galaxies with $z>1.4$ constitute roughly half 
of the failed targets \citep{freeman_etal09}.  
Moreover, spurious photo-z outliers persist even with techniques developed 
in conjunction with spectroscopic data that span the region 
of parameter space occupied by the photometric sample.  
As a nearly contemporaneous example, \citet{nishizawa_etal10} construct 
galaxies using simple spectral templates assuming a number of 
particular stellar populations based on the COSMOS galaxy catalog.   
 Applying LePhare\footnote{http://www.cfht.hawaii.edu/~arnouts/LEPHARE/cfht\_lephare/lephare.html}
 to their mock spectra gives offset islands in $z-\zphot$ space containing more than $5\%$ 
of the probability, which remains true even after refining their 
redshift estimator.  
In the analysis of the COSMOS data, \citet{ilbert_etal09} achieve an outlier rate of $0.7\%$
for a subsample of their brightest objects $(17.5<i_{\mathrm{AB}}<22.5).$  
However, their outlier rate dramatically increases to $15.3\%$ when they apply their
photo-z techniques to a subsample of faint objects
$(22.5<i_{\mathrm{AB}}<24)$.
An alternative approach to photo-z calibration is adopted in 
\citet{cunha_etal09}, who applied their weighted training set method to 
both simulated and actual SDSS data.  Their methods substantially improve 
upon the ability to directly reconstruct the redshift distribution of a photometric sample, 
but errors in the reconstructed $\mathrm{N}(\mathrm{z})$ remain at the percent level.  
Thus, while contemporary photo-z codes do provide useful guidance, 
outlier fractions greater than $\sim 10^{-3}$ persist and can 
affect the dark energy program.  
The ability to either limit, or understand, such outlier populations significantly 
better than the current state-of-the-art will be necessary to exploit fully 
the promise of cosmic shear tomography.

\vspace*{18pt}
\acknowledgments

We would like to thank Carlos Cunha, Joe Davola, Scott Dodelson, Salman Habib, 
Wayne Hu, Zeljko Ivezic, Bhuvnesh Jain, Arthur Kosowsky, Dan Matthews, Jeff Newman, 
Hiro Oyaizu, Martin White, and Michael Wood-Vasey 
for useful discussions and email exchanges during the course of this work.  
We are particularly grateful to Gary Bernstein and Alexia Schulz for detailed 
comments on an early draft of this manuscript.  
APH and ARZ are supported by the University of Pittsburgh and by the National 
Science Foundation (NSF) through grant NSF AST 0806367 and by the Department 
of Energy (DOE).  ZM is supported by the DOE under contracts DOE-DE-AC02-98CH10886 and 
DOE-DE-FG02-95ER40893.  DH is supported by the DOE OJI under contract 
DOE-DE-FG02-95ER40899, NSF under grant NSF AST 0807564, and the 
National Aeronautics and Space Administration under grant NNX09AC89G.  
ARZ and APH thank the organizers of the 2009 Santa Fe Cosmology 
Workshop, where a significant portion of this work was completed.  
ARZ thanks the Michigan Center for Theoretical Physics at the University of Michigan 
for hospitality and support while some of this work was performed.  
This research made use of the National Aeronautics and Space 
Administration Astrophysics Data System.

\begin{appendix}

\section{The Nonlinear Power Spectrum and Photometric Redshift Calibration Requirements}
\label{section:nlpower}

Much of the constraining power of weak lensing surveys arises from measurements on scales where 
the structures causing the deflections are undergoing nonlinear gravitational evolution \citep[e.g.][]{huterer_takada05}.  
Restricting consideration to large scales significantly degrades cosmological constraints 
\citep[e.g.][]{huterer02,huterer_takada05,zentner_etal08,schmidt08,hearin_zentner09}, so it is necessary to model 
nonlinear evolution in order to utilize weak lensing to constrain dark energy.  At least three approximate and related 
techniques are in common use:  (1) the fitting formula of \citet{peacock_dodds96}, which is based on the 
HKLM method \citep{hamilton_etal91};  (2) the halo model \citep{scherrer_bertschinger91,seljak00,ma_fry00,scoccimarro_etal01,cooray_sheth02}; 
and (3) the fitting formula of \citet{smith_etal03}.  The works of \citet{ma_etal06} and \citet{ma_bernstein08} specifying requirements for 
photometric redshift calibration employ the \citet{peacock_dodds96} relation.

\begin{figure}[t]
\centering
\includegraphics[width=7.5cm]{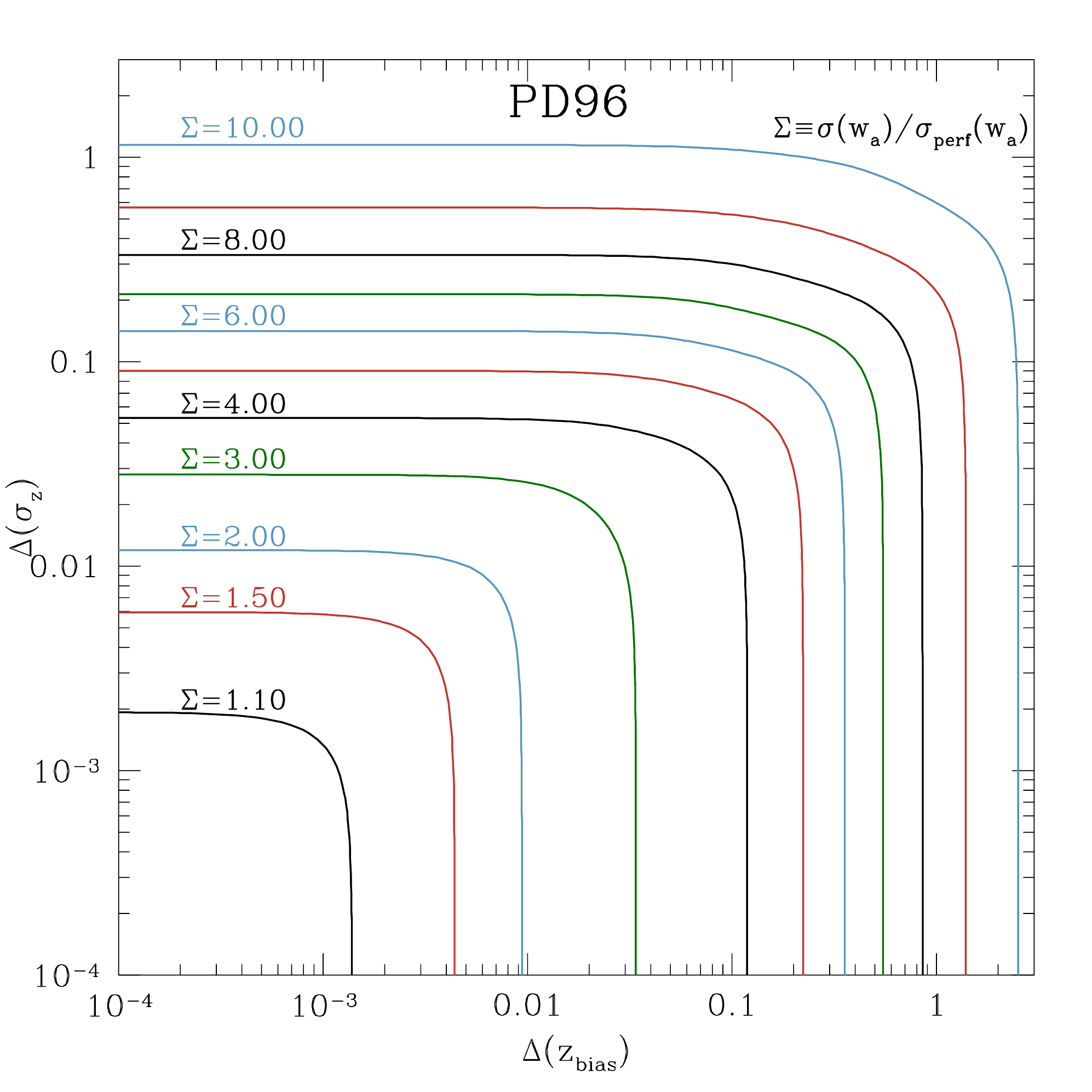}
\includegraphics[width=7.5cm]{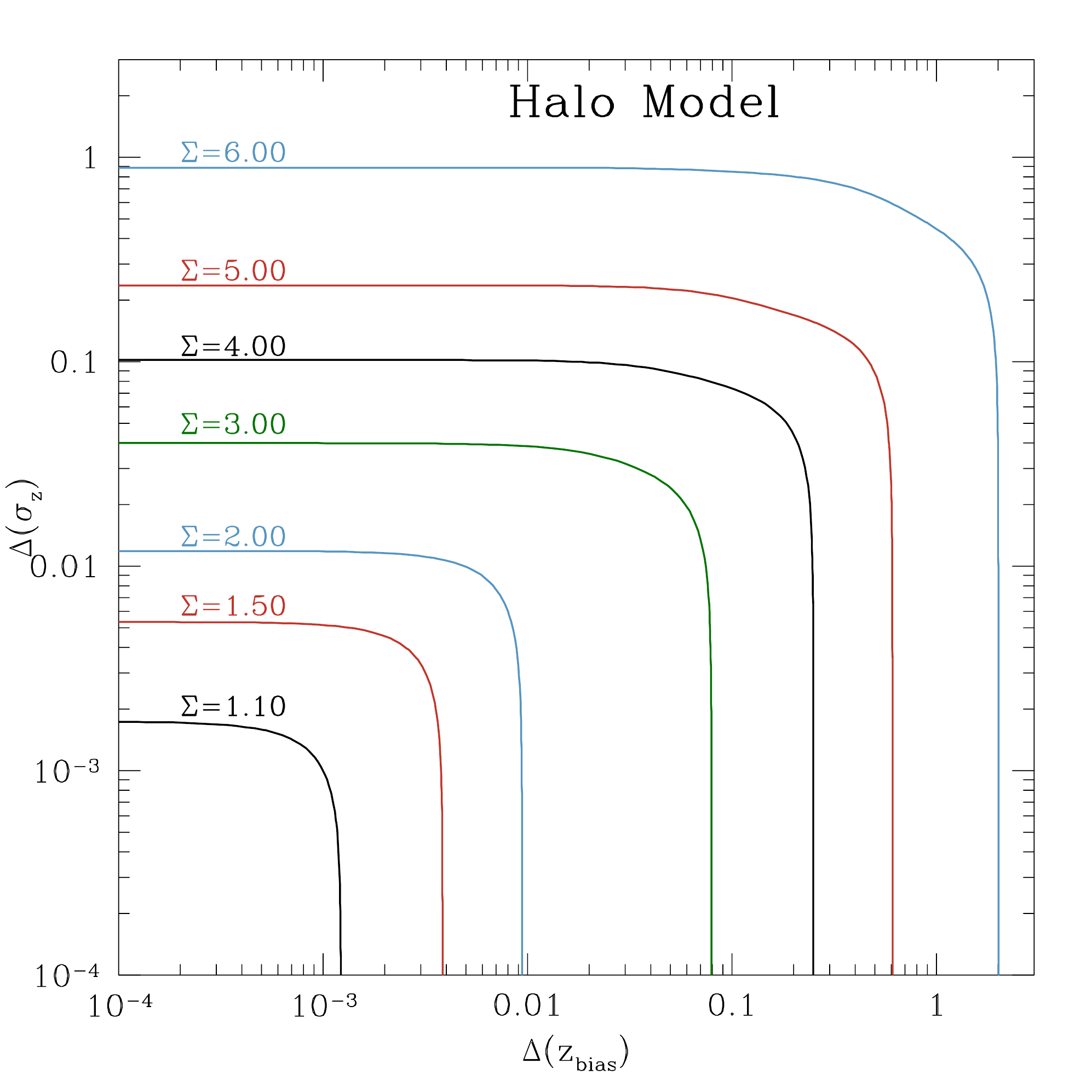}
\includegraphics[width=7.5cm]{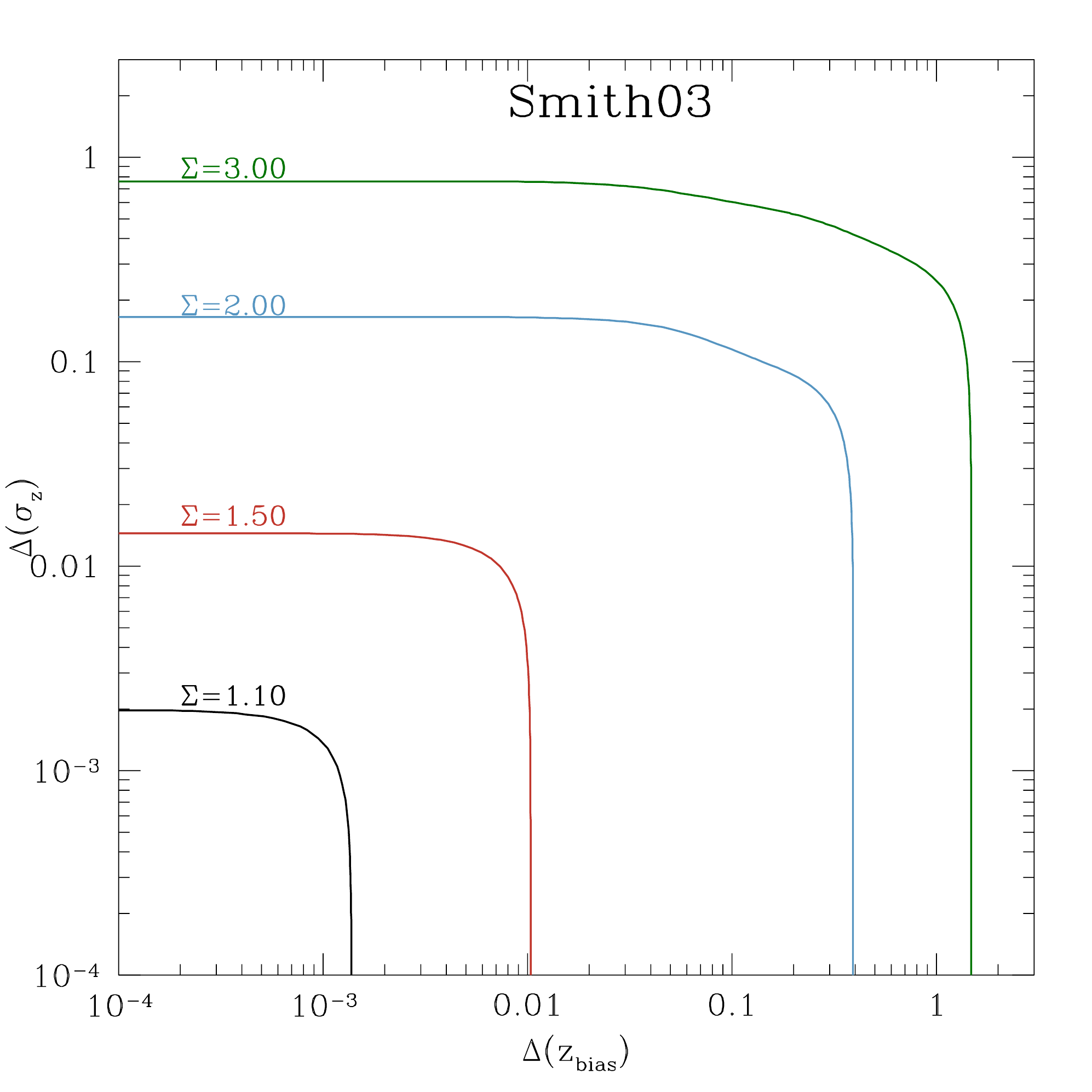}
\caption{
Contour plots for the level of $\wa$ constraint degradation as a function of 
priors on the photometric redshift scatter $\sigmaz$ and bias $\zbias$.  In this 
case, the priors are applied uniformly to the photometric redshift parameters 
are each redshift.  The contours demarcate equal parameter degradation defined as the error on 
$\wa$ after marginalizing over photometric reshift uncertainties.  We show constraints 
in units of the equivalant constraint in the limit of perfect knowledge of photometric redshift parameters, 
$\Sigma \equiv \sigma(\wa)/\sigma_{\mathrm{perf}}(\wa)$.  The {\em upper, left} panel was computed 
using the \citet{peacock_dodds96} fitting formula for the nonlinear power spectrum of 
density fluctuations and amounts to a near reproduction of the right panel of Figure~7 
in \citet{ma_etal06}.  The {\em upper, right} panel was computed using the halo model as described 
in \citet{zentner_etal08}.  The {\em bottom} panel was computed using the \citet{smith_etal03} relation 
for the nonlinear power spectrum of density fluctuations.  Significant differences between the 
levels of degradation are evident.  Note that in this figure, we use 
a different set of cosmological and experimental parameters so that this result is 
directly comparable to those in Figure~7 of \citet{ma_etal06}.
}
\label{fig:wacon}
\end{figure}

In the course of our study, we have recomputed the photometric redshift calibration requirements using each of the three approximate 
techniques mentioned in the previous paragraph.  In the limit of perfect knowledge of the photometric redshift distribution, 
each of these fitting formulas gives nearly identical dark energy constraints.  However, we have found 
that the photometric redshift calibration requirements have a strong dependence upon the 
method used to model nonlinear structure.  We summarize this finding in Figure~\ref{fig:wacon} where we display contours of constant 
degradation in the statistical error on $w_a$ as a function of both the prior on the bias $\Delta \zbias$ and the prior on the dispersion 
$\sigmaz$.  In other words, we show contours of $\sigma(\wa)$ in units of the statistical constraint on $\wa$ in the limit that the photo-z 
distribution parameters are known perfectly prior to the weak lensing analysis, $\sigma_{\mathrm{perf}}(\wa)$.  
We assume that the same priors are applied to all of our 31 dispersion parameters and 31 bias parameters at each 
redshift bin (see \S~\ref{sub:core}).  We summarize our findings in this way so that these results can be compared directly to 
Figure~7 in \citet{ma_etal06}.  To make the comparison as direct as possible, we have computed 
these forecasts using the fiducial cosmology and experimental setup of \citet{ma_etal06}, which differs 
slightly from those considered in the main text.  In this appendix only, our fiducial cosmology is 
$\omegam = 0.14$, $\omegab=0.024$, $\ns=1.0$, $\dr = 2.4 \times 10^{-5}$ (giving $\sigma_{8} \simeq 0.91$), 
and $\Omegade = 0.73$ combined with experimental parameters of $\fsky=0.1$ and $N^{A}=55\ \mathrm{arcmin}^{-2}$.

%

The upper left panel in Figure~\ref{fig:wacon} shows photo-z calibration requirements estimated using the \citet{peacock_dodds96} treatment of nonlinear 
power.  This panel shows nearly identical results to those in Figure~7 of \citep{ma_etal06} so that this panel validates our methods and 
provides a useful baseline to compare with the other panels.  According to this result, ensuring that constraints on $\wa$ are not degraded by 
more than a factor of two requires knowing the photo-z dispersion and bias to roughly $\sim 1\%$ prior to undertaking the weak lensing analysis.  
The upper, right panel of Fig.~\ref{fig:wacon} shows the same requirements constructed using the halo model for nonlinear clustering.  In the limit of 
restrictive prior knowledge of the photo-z distribution the \citet{peacock_dodds96} and halo model results yield nearly the same constraints.  When 
the photo-z distributions have relatively unrestrictive priors, the two techniques yield moderately different levels of projected degradation 
with, for example, uncertainty in the photo-z dispersion of $\Delta (\sigmaz) \approx 1$ corresponding to a factor of ten degradation in the 
\citet{peacock_dodds96} case but a factor of six degradation in the halo model calculation.

The largest differences among the forecasts comes from comparing the requirements using \citet{peacock_dodds96} to those computed using 
the \citet{smith_etal03} fit.  As with the halo model comparison, the different techniques agree well when prior knowledge of the photo-z distribution 
is very restrictive; thus as long as degradations due to photo-z uncertainty are $\lesssim10-20\%$ then it does not matter which technique one uses to predict the nonlinear evolution.  It is interesting that the constraints in the case of the halo model treatment degrade significantly less rapidly as prior knowledge becomes less and less restrictive.  Turning to the \citet{smith_etal03} fit, one would conclude that ensuring less than a factor of two degradation on the $\wa$ constraint requires $\sim 18\%$ knowledge of the dispersion and $\sim 40\%$ knowledge of the bias as compared to 
the $\sim 1\%$ requirements that result from the \citet{peacock_dodds96} analysis.

Clearly, at most one of these treatments can represent the growth of cosmic structure faithfully.  In the main text, we presented results 
using the \citet{smith_etal03} formula because these authors perform a detailed numerical study that finds the \citet{peacock_dodds96} 
and simple implementations of the halo model to be imprecise on scales relevant for cosmic shear cosmology.  
In the context of these fitting formulae, we find that \citet{smith_etal03} predicts greater 
power than \citet{peacock_dodds96} on scales most relevant to lensing ($0.1 \lesssim k/h\mathrm{Mpc}^{-1} \lesssim 10$), 
particularly at high redshift.  At this point, it is not possible to make a firm statement as to which approach is correct, but an exhaustive 
simulation program similar to that being carried out by \citet{heitmann_etal05,heitmann_etal08,heitmann_etal09} may be capable of 
providing a more definitive resolution in the case of dissipationless evolution.  Additional effort will be needed to treat any 
modifications induced by the baryonic component of the universe 
\citep{white04,zhan_knox04,jing_etal06,rudd_etal08,zentner_etal08,stanek_etal09,guillet_etal09}.

\end{appendix}

\bibliography{catphz}

\end{document}